\documentclass[prd,aps,floats,nofootinbib,showpacs]{revtex4}
\usepackage{epsfig,epsf}
\usepackage{amssymb}
\usepackage{amsmath}
\usepackage{graphicx}
\newcommand{\be}{\begin{equation}}
\newcommand{\ee}{\end{equation}}
\newcommand{\gsim}{\, \raisebox{-0.8ex}{$\stackrel{\textstyle >}{\sim}$ }}

\newcommand{\roughly}[1]%
{\mathrel{\raise.4ex\hbox{$#1$\kern-.75em\lower1ex\hbox{$\sim$}}}}

\newcommand\beq{\begin{eqnarray}}
\newcommand\eeq{\end{eqnarray}}

\def\Dsl{\,\raise.15ex \hbox{/}\mkern-12.8mu D}
\newcommand\Tr{{\rm Tr\,}}

\def\fm3{fm$^{-3}$}

\begin{document}
%\begin{frontmatter}
%
\preprint{\vbox{\hbox{LA-UR-04-3171}}}

\bigskip
\bigskip

\title{Neutrino scattering off pair-breaking and collective
excitations in superfluid neutron matter and in color-flavor locked
quark matter}

\author{Joydip Kundu$^1$ and Sanjay Reddy$^2$}

\affiliation{$^1$ Center for Theoretical Physics, Massachusetts
Institute of Technology, Cambridge, MA 02139 USA\\ $^2$ Theoretical
Division, Los Alamos National Laboratory, Los Alamos, NM 87545 USA}

\begin{abstract}
We calculate the correlation functions needed to describe the linear
response of superfluid matter, and go on to calculate the differential
cross section for neutral-current neutrino scattering in superfluid
neutron matter and in color-flavor locked quark matter (CFL). We
report the first calculation of scattering rates that includes
neutrino interactions with both pair-breaking excitations and
low-lying collective excitations (Goldstone modes). Our results apply
both above and below the critical temperature, allowing use in
simulations of neutrino transport in supernovae and neutron stars.
\end{abstract}
\maketitle

\section{Introduction}
The ground state of QCD at large densities is color superconducting
quark matter \cite{Barrois:1977xd, Bailin:1984bm,
Alford:1998zt,Rapp:1998zu}.  When the effects of quark masses can be neglected, 
three flavor quark matter will be in a particularly
symmetric phase called the Color-Flavor Locked (CFL) phase, with BCS
pairing involving all nine quarks \cite{Alford:1998mk}.  In this phase
the fermion excitation spectrum has a gap $\Delta$, and model
calculations indicate that $\Delta \approx 10 - 100$ MeV for quark chemical potential 
$\mu \approx
300 - 500$ MeV.  This phase breaks the ${SU(3)_{\rm color}} \times SU(3)_L
\times SU(3)_R \times U(1)_B$ symmetry of QCD down to the global
diagonal $SU(3)$ symmetry.  The lightest excitations are an octet of
pseudo-Goldstone bosons and a true Goldstone boson associated with the
breaking of the global $U(1)_B$ symmetry.  At some densities, 
however, the strange quark
mass may induce an appreciable stress on the symmetric CFL state, and less 
symmetric phases may be possible.  One possibility is the CFLK$^0$ phase,
which exhibits a Bose condensate of $K^0$ in addition to the diquark
condensate of the CFL phase--this phase breaks hypercharge and isospin
symmetries \cite{Bedaque:2001je, Kaplan:2001qk}.  Another possibility
is the LOFF phase, which exhibits crystalline color
superconductivity--the diquark condensate varies periodically in
space, breaking translation and rotation symmetries
\cite{Alford:2000ze, Bowers:2001ip, Leibovich:2001xr, Kundu:2001tt,
Casalbuoni:2001ha, Casalbuoni:2002hr,
Casalbuoni:2002my,Casalbuoni:2002pa}.  Most recently, a
superconducting phase of three-flavor quark matter with non-trivial
gapless fermionic excitations has been suggested \cite{Alford:2003fq}.
 
The densities at which color superconducting quark matter exists could
be attained in compact ``neutron'' stars or core-collapse supernovae.
It is therefore important to explore the impact of color
superconductivity on observable aspects of these astrophysical
phenomena.  Investigations to that end have included studies of
magnetic properties of neutron stars \cite{Alford:1999pb} and the
equation of state of dense matter \cite{Alford:2002rj, Baldo:2002ju,
Banik:2002kc,Shovkovy:2003ce,Buballa:2003et,Ruster:2003zh}.  The
interactions of neutrinos with superconducting quark matter has also
been explored \cite{Carter:2000xf, Jaikumar:2001hq, Reddy:2003ap}.
The emission of neutrinos from CFL during the long-term cooling epoch
was studied in Ref. \cite{Jaikumar:2002vg}.  During this epoch the
temperature is $T \lesssim 10^{10}$ K, and most of the excitations of 
CFL have small number densities.  As a result, neutrino emission
is highly suppressed.  Complementing Ref.~\cite{Jaikumar:2002vg} is
Ref.~\cite{Reddy:2002xc}, which studied the emission of neutrinos from
CFL in a young, proto-neutron star. During this epoch, the temperature
is $T \approx 10^{11}$ K, and the number densities of the excitations are
significant.  The authors of Ref.~\cite{Reddy:2002xc} only studied the interactions
of neutrinos with the Goldstone bosons, using the effective field
theory relevant at energies small compared to the gap.  But one might
expect the fermionic excitations to become relevant at $T \approx
10^{11}$ K, since these temperatures are close to the critical
temperature of CFL, and the gap will be suppressed.

These are the circumstances we analyze in this paper.  We study
neutral-current neutrino scattering in quark matter, above and
below the critical temperature for CFL.  We do this by calculating the
quark polarization tensor, including the effects of pair-breaking,
fermionic excitations.  We also include the effects of the collective,
bosonic excitation associated with the breaking of $U(1)_B$ by using
the Random Phase Approximation (RPA) to build this mode out of microscopic
quark-quark interactions.  This excitation has the largest
contribution of any of the bosonic modes.  We begin setting up the 
calculation in Section II.  But before proceeding to study CFL,
we consider
the case of non-relativistic fermionic modes, to make contact with
previous work and to motivate the use of RPA, arguing that consistency
with current conversation requires the inclusion of the collective
mode--we do this in Section III, where we also go on the calculate the
differential cross section for neutrino scattering in superfluid
neutron matter.  In Section IV we compute the medium polarization
tensor for a one component relativistic superfluid.  In Section V we
compute the medium polarization tensor for CFL, and go on to calculate
the differential cross section.  We conclude in Section VI with
reflections on this work.  Further details of the calculation for the
one-component, relativistic superfluid can be found in Appendix A.
Further details of the calculation for CFL can be found in Appendix B.

\section{Preliminaries}
The goal of this article is the calculation of the differential cross
section for neutral-current scattering of neutrinos in a dense quark
medium inside a proto-neutron star.  Neutrinos in a proto-neutron star
have typical energies $E_{\nu} \ll M_Z$, so we may write the neutrino
coupling to quarks as the four-Fermion effective lagrangian
\be 
{\cal L}_{Z} = \frac{G_F}{\sqrt{2}}~l^{\mu}~j_{\mu},
\label{lz1}
\ee 
where $l^\mu=\bar{\nu}~\gamma^\mu(1-\gamma_5)~\nu$ is the neutrino current,
and $j_\mu=\bar{q}~\gamma_\mu(c_V-c_A\gamma_5)~q$ is the quark weak neutral
current.  The differential cross section per unit volume in quark
matter can be expressed in terms of the quark current-current
correlation function, also called the polarization tensor, $\Pi_{\mu\nu}$
\cite{Horowitz:1991it,Reddy:1998hb}:
\begin{eqnarray}
\frac {1}{V} \frac {d^3\sigma}{d^2\Omega_3 dE_\nu'} &=& -\frac
{G_F^2}{32\pi^3} \frac{E_\nu'}{E_\nu}~
\frac{\left[1-n_{\nu}(E_\nu')\right]}{\left[1-\exp{\left(-q_0/T\right)}\right]}
~{\rm Im}~(L^{\mu\nu}\Pi_{\mu\nu}) \,,
\label{dcross}
\end{eqnarray}
where $E_\nu$ ($E_\nu'$) is the incoming (outgoing) neutrino energy,
$q_0=E_\nu-E_\nu'$ is the energy transferred to the medium, and $T$ is the
temperature of the medium. The factor $[1-\exp(-q_0/T)]^{-1}$ ensures 
detailed balance.  The factor $[1-n_{\nu}(E_\nu')]$, where
\begin{equation}
n(E) = \frac{1}{e^{(E-\mu)/T}+1},
\label{fermi}
\end{equation}
enforces the blocking of final states for the
outgoing neutrino.  The neutrino tensor $L^{\mu\nu}$ is given by
\begin{equation}
L^{\mu\nu}= 8[2k^{\mu}k^{\nu}+(k\cdot q)g^{\mu\nu}
-(k^{\mu}q^{\nu}+q^{\mu}k^{\nu})\mp i\epsilon^{\mu\nu\alpha\beta}
k_{\alpha}q_{\beta}] \,,
\end{equation}
where the incoming four-momentum is $k^\mu$ and the momentum
transferred to the medium is $q^\mu$.  The minus (plus) sign on the
final term applies to neutrino (anti-neutrino) scattering.  The
response of the medium is characterized by the polarization tensor,
$\Pi^{\mu\nu}$. In the case of free quarks,
\beq
\Pi^{\mu\nu}(Q)= -i  \int
\frac{d^4 P}{(2\pi)^4} {\rm Tr}~[S_0(P)\Gamma^\mu 
S_0(P+Q)\Gamma^\nu] \,, 
\label{pi_free}
\eeq
where $S_0(P)$ is the free quark propagator at finite temperature and
chemical potential, and $\Gamma^\mu = \gamma^\mu(c_V-c_A\gamma_5)$.
The inner trace is over Dirac, flavor, and color indices.  The free
quark propagator is \cite{Rischke:2000qz}
\beq
S_0(P) = \delta^{c}_{d}\delta^{f}_{g}~\left( \frac{p_0 + E_{\bf
p}}{p_0^2 - E_{\bf p}^2} \; \Lambda^+_{\bf p} \gamma_0 \; + \;
\frac{p_0 - \bar{E}_{\bf p}}{p_0^2 - \bar{E}_{\bf p}^2} \;
\Lambda^-_{\bf p} \gamma_0\right)\,,
\eeq
where ${c,d}$ and ${f,g}$ denote color and flavor indices,
respectively, and
\beq
\Lambda^{\pm}_{\bf p} = \frac{1}{2} ( 1 \pm \gamma_0 \vec{\gamma} \cdot {\bf \hat{p}})
\label{sfree}
\eeq
are the positive and negative energy projection operators. The
energies in the propagator are measured relative to the Fermi surface.
To wit, $E_{\bf p}=p-\mu$ for (massless) particle states, and
$\bar{E}_{\bf p} =p + \mu$ for anti-particle states.

Since we are interested in the interaction of neutrinos with a
superconducting medium, we introduce the Nambu-Gor'kov formalism
\cite{Gor'kov:1959,Nambu:1960tm}, which allows the incorporation of
diquark correlations into the quark propagator.  This is done by
artificially doubling the quark degrees of freedom by introducing
charge conjugate field operators $q_C$ and $\bar{q_C}$, defined by
$$ q = C \bar{q_C}^T \qquad\mbox{and}\qquad \bar{q} = q_C^T C,$$ where
$C = i \gamma_0 \gamma_2$ is the charge conjugation matrix (with $C =
-C^{-1} = -C^T = -C^\dagger$ and $C \gamma_\mu C^{-1} =
-\gamma_\mu^T$). 
The Nambu-Gor'kov field is given by 
\beq
\Psi \equiv \left( \begin{array}{c} q \\ q_C \end{array} \right)
\qquad\mbox{and}\qquad \bar{\Psi} \equiv \left(\begin{array}{cc}
\bar{q} & \bar{q_C} \end{array} \right).
\label{nambufields} 
\eeq
In terms of these fields, the weak interaction Lagrangian,
Eq.~(\ref{lz1}), becomes
\begin{equation} 
{\mathcal L}_Z = \frac{G_F}{\sqrt{2}} \; \bar{\nu} \gamma_\mu (1-\gamma_5) 
\nu \; \bar{\Psi} \Gamma_Z^\mu   \Psi 
\label{lz2a}
\end{equation}
where the neutrino-quark vertex in the Nambu-Gor'kov space is 
\begin{equation}
\Gamma_Z^\mu=\left( \begin{array}{cc}  \gamma^\mu (c_V - c_A \gamma_5)
& 0 \\ 0 &  -\gamma^\mu (c_V + c_A \gamma_5) \end{array} \right) \,.
\label{lz2b}
\end{equation}
The Nambu-Gor'kov propagator that includes BCS diquark correlations in
the mean-field approximation is given by
\cite{Pisarski:1999av,Rischke:2000qz, Rischke:2000ra}
\be
S(P)  = \left(
\begin{array}{cc} 
G^+(P) & \Xi^-(P) \\ \Xi^+(P) & G^-(P) \end{array}
\right)\,,
\label{NGprop}
\ee
where
$$
G^+(P) = \frac{p_0 + E_{\bf p}}{p_0^2 - \xi^2_{\bf p}}~
\Lambda^+_{\bf p} \gamma_0,  \qquad 
G^-(P) = \frac{p_0 - E_{\bf p}}{p_0^2 - \xi^2_{\bf p}} ~
\Lambda^-_{\bf p} \gamma_0, $$ 
\beq \Xi^+(P) = \frac{-\Delta}{p_0^2 - \xi^2_{\bf p}} ~\gamma_5
\Lambda^-_{\bf p}, \qquad \Xi^-(P) = \frac{\Delta}{p_0^2 - \xi^2_{\bf p}}~
\gamma_5 \Lambda^+_{\bf p} \,,
\eeq
and
$$ \xi_{\bf p} = \sqrt{E_{\bf p}^2 + \Delta^2}.$$
We have suppressed any color-flavor indices, and we are neglecting the contribution of antiparticles.

\section{A non-relativistic detour} 

Although we will ultimately be interested in relativistic,
superconducting quark matter, we digress to consider a one component
system of non-relativistic superfluid baryons, such as neutron matter
with $^1S_0$ pairing, which is itself relevant to neutrino transport
in neutron stars and supernovae.  This will allow us to make contact
with the vast body of published work on superconductivity in
non-relativistic systems and will serve as a pedagogical prelude to
the relativistic case.

For non-relativistic fermions the structure of the Nambu-Gor'kov
propagator greatly simplifies and is given by a $2\times 2$ matrix
\cite{Schrieffer:1964}.  We shall consider a simple model for the
pairing interaction. In this case the non-relativistic Hamiltonian, in
terms of the fermion creation and annihilation operators, is \be
\mathcal{H} -  \mu N = \sum_p~ E_p ~a^\dag_{\vec{p}}~ a_{\vec{p}} ~+~ G \sum_p
a^\dag_{\vec{p}}a^\dag_{-\vec{p}}a_{-\vec{p}}a_{\vec{p}}\,,
\label{nrlagrangian}
\ee where $N= \sum a^\dag a$ is the number operator, $E_p=p^2/2M
-\mu$, $M$ is the mass and $\mu$ is the non-relativistic chemical
potential. In the mean field or BCS approximation, we replace the
pair operator by the space and time independent classical expectation
value. This defines the constant field $\Delta^\dag=-G\sum_p
a^\dag_{\vec{p}}a^\dag_{-\vec{p}}$ and $\Delta=-G\sum_p
a_{\vec{p}}a_{-\vec{p}}$. The solution of the gap equation, namely
$\partial(\mathcal{H}- \mu N )/\partial\Delta=0$ determines the
magnitude of $\Delta$.  The propagator is given by
\be 
S(P) =
\left( \begin{array}{cc} G^+(P) & \Xi^-(P) \\ \Xi^+(P) & G^-(P)
\end{array} \right) \,,
\label{NRNGprop}
\ee
where
$$
 G^+(P) = \frac{p_0 + E_{\bf p}}{p_0^2 - \xi^2_{\bf p}}\,, \qquad
G^-(P) = \frac{p_0 - E_{\bf p}}{p_0^2 - \xi^2_{\bf p}} $$
\beq
\Xi^+(P) = \frac{\Delta}{p_0^2 - \xi^2_{\bf p}}\, \qquad
\Xi^-(P) = \frac{\Delta^\dag}{p_0^2 - \xi^2_{\bf p}} \,.
\eeq
The quasi-particle energy is given by
\begin{eqnarray}
\xi_{\bf p} &=&
\sqrt{E_{\bf p}^2 + \Delta^2}\,,\quad
\mathrm{where }\quad E_{\bf p} = \frac{p^2}{2 M} - \mu \nonumber \,.
\end{eqnarray}

The neutrinos couple to the weak neutral-current of the neutrons,
which has a contribution from both the vector current and the axial
vector baryon current (see Eq.~(\ref{lz1})).  The neutral-current
carried by the neutrons is given by
$$ j_\mu = \Sigma_{\vec{p}}~\Psi^\dagger(p)~\left(c_V~\gamma_\mu
(p+q,p)~+~c_A~\gamma^A_\mu (p+q,p)\right)~\Psi(p+q) \,,
\label{vector_current}
$$ where $c_V=1/2$ and $c_A=1.23/2$. In the non-relativistic limit
$\gamma_\mu(p+q,p)=(\tau_3,0)$ and
$\gamma^A_\mu(p+q,p)=(0,\vec{\sigma}~\hat{1} )$, with
$\tau_3=\mathrm{diag}(1,-1)$ and $\hat{1}=\mathrm{diag}(1,1)$ in
Nambu-Gor'kov space, and $\vec{\sigma}$ is the Pauli matrix in spin space. To begin, we focus on the vector current, though we
will return to the axial vector current.

Conservation of the vector baryon current means that the vector
response function
\be 
\Pi_{\mu \nu}^V(q_0,\vec{q}) \equiv ~-\imath \int~\frac{d^4p}{(2\pi)^4}  
~\Tr [S(p)~\gamma_\mu(p,p+q)~S(p+q)~\gamma_\nu (p+q,p)]
\label{nrpi00}
\ee
must satisfy the conservation equation
\beq
q^{\mu}~\Pi_{\mu \nu} = \Pi_{\mu \nu}~q^{\nu}=0\,.
\label{currentconserve}
\eeq 
The polarization function in Eq.~(\ref{nrpi00}) does not satisfy
this equation.  This violation can be traced to the use of the {\it
dressed} mean-field Nambu-Gor'kov propagator on the one hand, and the
use of the {\it bare} vertex on the other.  To ensure local baryon
conservation, we must use a dressed vertex, $\tilde{\Gamma}(p+q,p)$,
which must satisfy the generalized Ward identity for the superfluid:
\be q^{\mu} \tilde{\Gamma}_{\mu}(p+q,p) = \tau_3 S^{-1}(p) -
S^{-1}(p+q) \tau_3 \,.
\label{GWI}
\ee That the bare vertex $\gamma_\mu(p+q,p)$ is not sufficient to
satisfy current conservation was realized shortly after the
development of the theory of superconductivity by Bardeen, Cooper and
Schrieffer \cite{Bardeen:1957mv} in independent articles by Bardeen
\cite{Bardeen}, Bogoliubov \cite{Bogoliubov:1,Bogoliubov:2}, Anderson
\cite{Anderson:1958pb}, and Nambu \cite{Nambu:1960tm}.  It is this
correction that naturally leads us to incorporate the Goldstone mode.

\begin{figure}[t]
\begin{center}
\includegraphics[width=11cm]{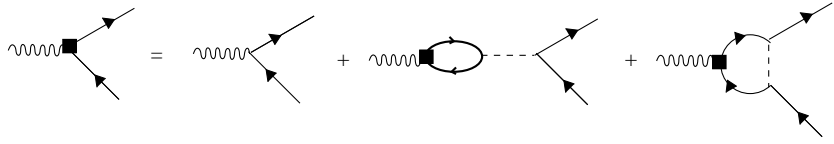}
%\centerline{\epsfxsize 11cm \epsffile{vertex1.eps}}
\end{center}
\caption{Feynman diagram corresponding to the vertex equation
(Eq.\ref{vertexeqn}). The squares represent the dressed weak
interaction vertex, and dashed lines represent the interaction between
the fermions. The solid, directed lines represent Nambu-Gorkov propagators.}
\label{fig_vertex}
\end{figure}

The dressed vertex $\tilde{\Gamma}$ compatible with Eq.~(\ref{GWI})
satisfies the following integral equation \be \tilde{\Gamma}_\mu
(p+q,p) = \gamma_{\mu} (p+q,p) + G~\imath \int \frac{d^4p'}{(2 \pi)^4}
~ \tau_3 S(p'+q) \tilde{\Gamma}_\mu (p'+q,p')S(p')\tau_3~ \,,
\label{vertexeqn}
\ee where $G$ is the 2-body contact interaction defined by the
Hamiltonian in Eq.~(\ref{nrlagrangian}). The behavior of the dressed
vertex function in the long wavelength limit can be inferred from the
Ward identity in Eq.~(\ref{GWI}). The right-hand side of Eq.~(\ref{GWI}), in
the $q\rightarrow 0$ limit, reduces to \be \lim_{\vec{q}\rightarrow 0}
~\left( \tau_3S^{-1}(p)-S^{-1}(p+q)\tau_3 \right) = 2 \imath~ \Delta
\tau_2 \,.  \ee The Ward identity therefore dictates that the vertex
function be singular for zero energy transfer ($ q_0=0$) when the
three momentum transfer vanishes ($\vec{q} \rightarrow 0$). This
singularity is characteristic of the presence of a zero energy
collective mode -- the Goldstone mode expected on general grounds
\cite{Goldstone:1961eq}. One may obtain the dispersion relation for
the Goldstone mode by inspecting the pole structure of the solutions
of Eq.~(\ref{vertexeqn}).  For the simplified interaction considered
here, the integral equation for the vertex function is represented by
the Feynman diagrams shown in Fig.~\ref{fig_vertex}.

Calculating the vertex function by retaining only the terms shown in
the figure is equivalent to the Random Phase Approximation (RPA)
\cite{Schrieffer:1964}. Note that it contains both the direct and
exchange diagrams (the second and third terms, respectively, on the
right-hand side of Fig.~\ref{fig_vertex}).  
We only need to solve for $\tilde{\Gamma}_0$, rather
than all four components of $\tilde{\Gamma}_\mu$, since for small
$\vec{q}$ the dominant response of non-relativistic neutrons is due to
the coupling of the baryon density to neutrinos; the coupling of the
velocity-dependent components of the baryon current to neutrinos is
suppressed by a factor $q/M$ and may be neglected.  An analytic
solution for $\tilde{\Gamma}_0$ exists and is given by $$
\tilde{\Gamma}_0 = \frac{1}{\chi_0}~\tau_3 + ~\frac{1}{\chi}~\gamma_H
\, \quad\quad \mathrm{with}\quad \gamma_H = \left( \begin{array}{cc}
0& N_1 \\ N_2&0 \end{array} \right)\,,
$$
where 
\begin{eqnarray*} 
N_1 &=& 2\left(\Pi_{11:12} + \Pi_{11:12}\Pi_{22:11} -
\Pi_{12:11}\Pi_{12:12}\right) \,, \\ N_2 &=& 2\left(\Pi_{12:11} +
\Pi_{12:11}\Pi_{11:22} - \Pi_{11:12}\Pi_{12:12} \right)\,, \\ \chi_0
&=& (1 + \Pi_{12:12} - \Pi_{11:11})\,, \\ \mathcal{D} &=& 1 -
\Pi_{12:12}^2 + \Pi_{11:22} + \Pi_{22:11} + \Pi_{11:22}\Pi_{22:11} \,,
\\ \chi&=& 4 \Pi_{11:12} \Pi_{12:11} \Pi_{12:12} - \left( 1 -
\Pi_{12:12} \right) \Pi_{12:12}^2 \\ &-& \left( 1 + \Pi_{11:22}
\right) \left( -1 + 2 \Pi_{12:11}^2 + \Pi_{12:12} \right) +
\Pi_{22:11} \\ &-& \left( \Pi_{11:22} \left( -1 + \Pi_{12:12} \right)
+ \Pi_{12:12} \right) \Pi_{22:11} \\ &-& 2 \Pi_{11:12}^2 \left( 1 +
\Pi_{22:11} \right) + \Pi_{11:11} \left( 1 - \Pi_{12:12}^2 +
\Pi_{22:11} + \Pi_{11:22} \left( 1 + \Pi_{22:11} \right) \right) \\
\Pi_{ij:kl} &=&~G~\imath \int\frac{d^4p}{(2\pi)^4} ~
S_{ij}(p+q)~S_{kl}(p) \,,
\end{eqnarray*}
and $S(p)$ is the mean field propagator appearing in
Eq.~(\ref{NRNGprop}).  The complex function $\chi(q_0,q)=0$ when
$q_0=q~c_s,$ where $c_s=p_F/(\sqrt{3}M)$ is the sound velocity of
the neutron gas (for $q\ll p_F$). The vertex is singular at this
point, indicating the presence of the Goldstone mode in the response.
Note that the singularity does not appear in the vertex proportional
to $\tau_3$ in NG space. The off-diagonal nature of the singularity is
characteristic of fluctuations of the phase of $\Delta$ - which is
the Goldstone excitation.  Further, at small energy transfer $q_0 \ll
\Delta$ the expression for $\chi$ simplifies and we find that
$\chi\simeq \chi_0 \mathcal{D}$. The Goldstone singularity occurs when
$\mathcal{D}=0$ and $\chi_0\neq 1$ is a result of additional
screening corrections included in RPA in both the quasi-free and
Goldstone-mode responses.

\begin{figure}[]
\begin{center}
\includegraphics[width=16cm]{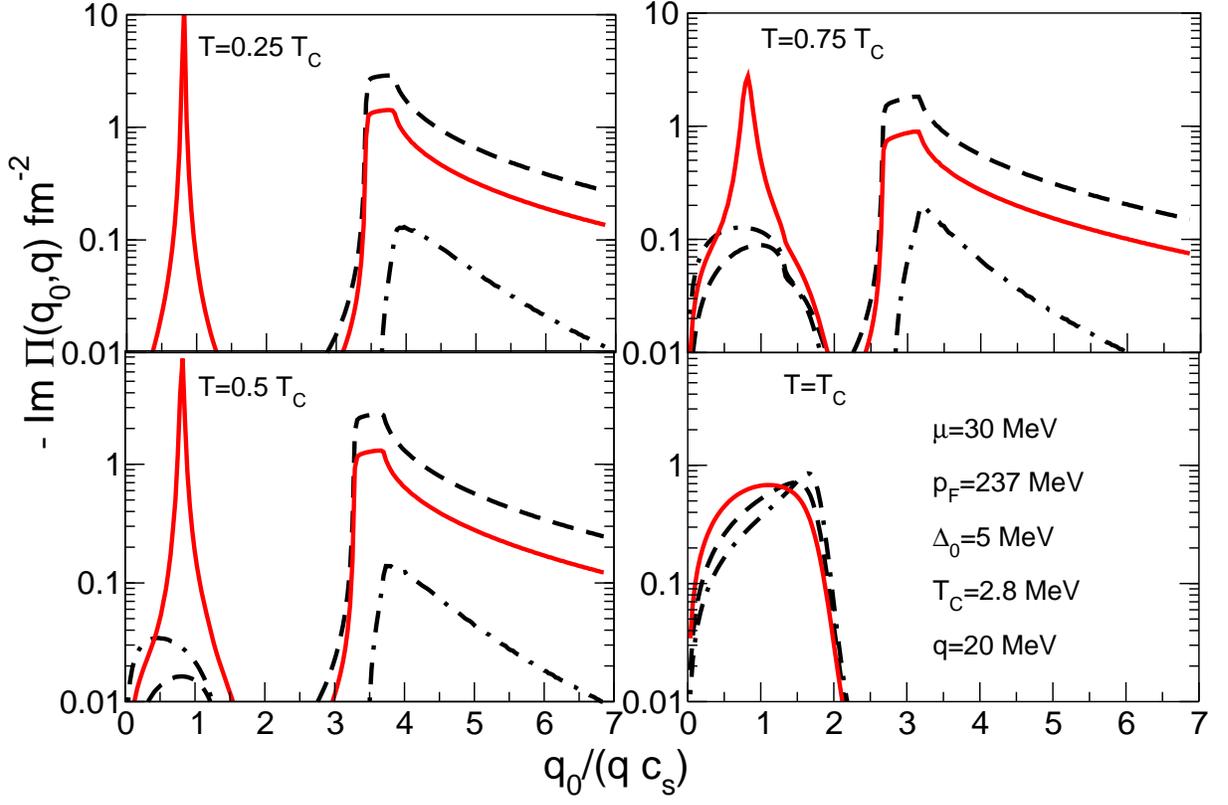}
%\centerline{\epsfxsize 14cm \epsffile{impi_mu30delz5q20.eps}}
\end{center}
\caption{We plot -Im$\Pi$ as a function of the dimensionless variable
$q_0/(q c_s)$, where $c_s = p_F/(\sqrt{3} M)$ and with fixed momentum
transfer $q=0.1 p_F$. The dashed line is the quasi-free vector
response, -Im$\Pi^{qf}_V$, and the solid line is the full RPA (that 
is, quasi-free plus collective with screening corrections) vector response,
-Im$\Pi^{RPA}_V$.  The spike at $q_0/(q c_s)\simeq1$ is the Goldstone
mode response. The dot-dashed line is the RPA axial response,
-Im$\Pi^{RPA}_A$. We use neutron chemical potential $\mu =
p_F^2/(2M)=30$ MeV.  The gap at zero temperature is $\Delta = 5$ MeV,
and the critical temperature is $T_c = 2.8$ MeV.}
\label{nrimpiz}
\end{figure}

In RPA,  the polarization tensor which satisfies the Ward identity is 
\be 
\Pi_V^{RPA}(q_0,\vec{q})= ~-\imath \int~\frac{d^4p}{(2\pi)^4}  
~\Tr [S(p)~\tilde{\Gamma}_0~S(p+q)~\tau_3]
\label{nrpirpa}
\ee
The simple form of the solution for the dressed vertex for $q_0
\ll \epsilon_F$ allows us to write
\begin{equation}
\Pi_V^{RPA}(q_0,\vec{q}) =~\frac{1}{\chi_0}~\Pi_V^{qf}(q_0,\vec{q}) +
\Pi^H(q_0,\vec{q}) \,,\\
\label{pi_rpa_nr}
\end{equation}
where
\begin{equation}
\Pi_V^{qf} (q_0,\vec{q}) = ~-\imath~\int~\frac{d^4p}{(2\pi)^4}  
~\Tr [S(p)~\tau_3~S(p+q)~\tau_3]\,
\end{equation}
and
\begin{equation}
\Pi^H(q_0,\vec{q}) =
~-\frac{\imath}{\chi_0~\mathcal{D}}~\int~\frac{d^4p}{(2\pi)^4} ~\Tr
[S(p)~\tau_3~S(p+q)~\gamma_H]\,.
\end{equation}
$\Pi^{qf}$ is just the quasi-free or mean-field response, and $\Pi_H$
is the response due to the Goldstone mode. We plot -Im$\Pi^{qf}_V$ and 
-Im$\Pi^{RPA}_V$ in Fig.~\ref{nrimpiz}, as the dashed line and the solid line, respectively, at various temperatures and ambient conditions
appropriate for the neutron superfluid in neutron stars, namely, for
chemical potential $\mu=p_F^2/2M=30$ MeV, and gap $\Delta=5$ MeV,
where $M$ and $p_F$ are the mass and Fermi momentum,
respectively, of neutrons.  Note that the quasi-free response has
support only for $q_0 \gsim 2\Delta$ at $T \ll \Delta$, due to the gap
in the fermion excitation spectrum.  Eq.~(\ref{pi_rpa_nr}) has $\chi_0
\neq 1$ because correlations result in screening of the coupling of
quasi-free excitations to the external current. In the weak-coupling
limit ($\Delta \lll \mu$), this correction is small, but it becomes
important even at moderate coupling strength, as seen in the figure.
At small but non-zero temperature ($T\ll \Delta$),
$\mathrm{Im}~\mathcal{D}(q_0,q)\neq 0 \sim \exp(-T/\Delta)$ due to a
small contribution from thermal quasi-particle excitations. This
reduces the singular behavior to a resonant response (Lorentzian) with
a small but finite width.  With increasing temperature, the collective
mode gets damped as the single-pair excitations become increasingly
important. This trend is clearly seen in the Fig.~\ref{nrimpiz}.
Finally at $T_c$, both the pair-breaking and Goldstone mode
excitations disappear and the single-pair response dominates. The
slight enhancement seen at small $q_0$ is a characteristic of
attractive correlations induced by the strong interactions in the
normal phase \cite{Reddy:1998hb}.

The response in the axial vector channel can also be calculated using
the method described above. In this case, the vertex does not exhibit
any singular behavior, since there is no Goldstone mode in this
channel. In a homogeneous and isotropic system, the axial polarization
tensor is diagonal, with equal components,
$\Pi^A_{11}=\Pi^A_{22}=\Pi^A_{33}=\Pi_A$. An explicit calculation
shows that the axial polarization tensor in RPA is
\begin{equation}
\Pi_A^{RPA}(q_0,\vec{q}) =~\frac{1}{\chi_0}~\Pi_A^{qf}(q_0,\vec{q})
\label{pia_rpa_nr}
\end{equation}
where
\begin{equation}
\Pi_A^{qf}(q_0,\vec{q}) = ~-\imath~\int~\frac{d^4p}{(2\pi)^4}  
~\Tr [S(p)~\hat{1}~S(p+q)~\hat{1}]\,. 
\end{equation}
The dot-dashed lines in Fig.~\ref{nrimpiz} show the behavior of the
imaginary part of the $\Pi_A^{RPA}(q_0,\vec{q})$ at fixed q as a
function of $q_0$. 
\begin{figure}[h]
\begin{center}
\includegraphics[width=14cm]{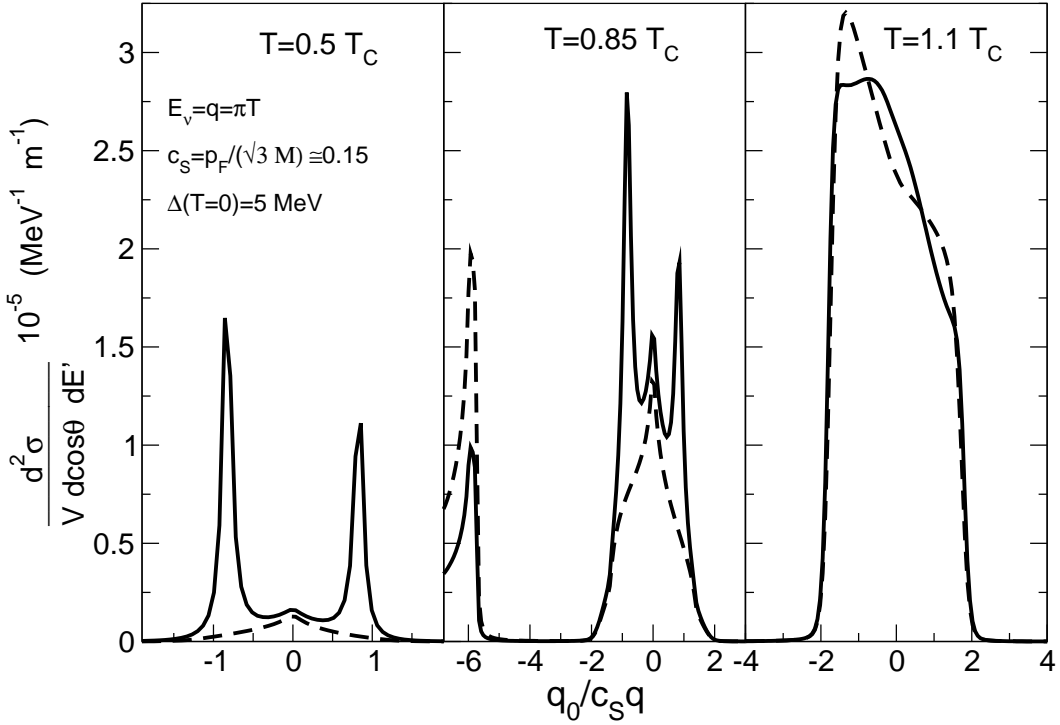}
%\centerline{\epsfxsize 14cm \epsffile{dsigma_nrdel5mu30.eps}}
\end{center}
\caption{We plot the differential cross section for neutral-current
neutrino scattering in superfluid neutron matter as a function of the
dimensionless variable $q_0/(q c_s)$, where $c_s = p_F/(\sqrt{3}
M)\simeq 0.15$. The solid curves are the RPA results which include the
Goldstone contribution and the dashed lines are the mean-field results.  
The incoming neutrino energy and the momentum transfer
$q$ were set equal to $\pi\times T$ - typical of thermal neutrinos.
The spike at $q_0/(q c_s)\simeq1$ is the Goldstone mode response. We
use neutron chemical potential $\mu = p_F^2/(2M)=30$ MeV. The gap at
zero temperature is $\Delta = 5$ MeV, and the critical temperature is
$T_c = 2.8$ MeV.}
\label{nrdiffcs}
\end{figure}
The differential cross section for neutrino scattering in the neutron
matter can be written in terms of the imaginary part of the
polarization tensor. The differential cross section per unit volume
for a neutrino with energy $E_\nu$ to a state with energy
$E'_{\nu}=E_\nu-q_0$ and scattering angle $\theta$ is given by
\begin{eqnarray}
\frac{d^2\sigma}{V~d\cos\theta ~dE'_\nu}
&=&~-\frac{G^2_{\mathrm{F}}}{4\pi^3}~{E'_\nu}^2 ~
\frac{1-n_\nu(E_\nu')}{1-\exp(-q_0 / T)} \nonumber \\ 
&\times&\left[c_V^2~(1+\cos\theta) ~\mathrm{Im}~\Pi^{RPA}_V(q_0,q) + 
c_A^2~(3-\cos\theta) ~\mathrm{Im}~\Pi^{RPA}_A(q_0,q) \right] \,.
\end{eqnarray}
These expressions permit calculation of the neutrino opacity in
superfluid neutron matter at arbitrary temperature. The results are
shown in Fig.~\ref{nrdiffcs}. The incoming neutrino energy $E_\nu$ and
the momentum transfer $q$ were set equal to $\pi T$. This is typical
for thermal neutrinos.  Note that the Goldstone mode continues to play
a role even when $T \simeq T_c/2$. Note also that with increasing
temperature the gap equation yields smaller gaps - seen in the
decreasing threshold for the quasi-free response. Although it is easy
to deduce the relative importance of RPA corrections to the response
from Fig.~\ref{nrdiffcs}, we present a table to quantify the
differences.  The table below provides a comparison between the differential
cross sections integrated over the energy transfer (area under the
curves shown in Fig.~\ref{nrdiffcs}). 
\begin{center}
\begin{tabular}{|c|c|c|} \hline
T & $d\sigma_{qf}/V dcos\theta~\times~10^{-5}$  m$^{-1}$ &
$d\sigma_{RPA}/V dcos\theta ~\times~10^{-5}$  m$^{-1}$ \\ \hline 
0.5 T$_C$ & 0.1 & 0.5 \\ \hline 
0.85 T$_C$ & 2.6 & 4.9 \\ \hline 
1.1 T$_C$ & 12.8 & 12.9 \\ \hline
\end{tabular}
\end{center}

Although scattering kinematics do not probe the response in the
time-like region, neutrino pair production does arise from time-like
fluctuations. As a result these same polarization functions can be
used in calculations of the neutrino emissivities in superfluid matter
- a process that is commonly referred to as the pair-breaking process
\cite{Flowers:1976}.

At low temperature and low energy, the response is dominated by the
Goldstone mode. In this regime it is
appropriate to use the low energy effective theory involving only the
Goldstone mode. The effective Lagrangian for the U(1) mode is given by 
\begin{equation} 
{\cal L}_{GB} = \frac{f_H^2}{8} (\partial_0 U \partial_0 U^{\dagger} - c_s^2 ~
 \partial_i U \partial_i U^{\dagger}) 
\end{equation} 
where $U=exp(2  \imath H/f_H)$, $H$ is the Goldstone field, and
$f_H^2=M~p_F/\pi^2$ is a low energy constant which is the equivalent
of the pion decay constant in the chiral lagrangian. We can compute
the coupling of the $H$ mode to the neutrinos by 
matching to the weak current in the microscopic theory.  We find
that the amplitude for the process $\nu \rightarrow H~\nu $ is given
by \cite{Bedaque:2003wj}
\begin{equation}
A_{\nu\rightarrow H \nu} = 
\frac{G_F}{\sqrt{2}}~c_V~f_H~\partial_0 H ~\bar{\nu}\gamma^0 \nu\,.
\label{nreftweak}
\end {equation} 
Using Eq.~(\ref{nreftweak}), we can compute the differential cross
section for neutrino scattering due to the ``Cerenkov'' process $\nu
\rightarrow H \nu$ \cite{Reddy:2002xc}.  We find that
\begin{equation} 
\frac{d^2\sigma}{V~d\cos\theta ~dE'_\nu}
=~-\frac{G^2_{\mathrm{F}}}{4\pi^2}~f_H^2~c_V^2~{E'_\nu}^2 ~
\frac{1-n_\nu(E'_\nu)}{1-\exp(-q_0 / T)}(1+\cos\theta) ~q_0~
\delta(q_0 -c_s q) \,.
\label{nreft}
\end{equation} 

The low energy and low temperature limit of the RPA response should agree
with the above result obtained using the effective theory. We show
that this indeed the case. We begin by noting that the RPA vertex
when $q_0 \ll \Delta$ and $q \ll k_F$ can be written as
\cite{Nambu:1960tm}
\begin{equation} 
\tilde{\Gamma_0}= \frac{2 \Delta q_0}{q_0^2 - c_s^2 q^2 }~i \tau_2 \,.
\end{equation} 
Substituting this result in Eq.~(\ref{pi_rpa_nr}) we find
for $q_0 \ge 0$ 
\begin{equation}
\mathrm{Im}~\Pi_V^{RPA}(q_0,q) = -\frac{M p_F}{2 \pi} q_0 ~\delta(q_0
- c_s q) \,.
\label{lowelowt}
\end{equation} 
The quasi-free response is exponentially suppressed at low temperature,
and using Eq.~(\ref{lowelowt}) it is easily verified that the
differential cross section in RPA agrees with Eq.~(\ref{nreft}). 
\section{Response of a One-Component, Relativistic Superfluid}
In the last section we argued that a consistent calculation of the
medium polarization tensor must include not only the quasi-free
response of the medium associated with the pair-breaking excitations,
but also the collective response associated with the massless
excitation $H$ arising from spontaneous breakdown of $U(1)_B$.  We
then calculated the medium polarization tensor for a one-component,
non-relativistic superfluid -- superfluid neutron matter.  In this section we will calculate the
medium polarization tensor for a one-component ultra-relativistic
superfluid -- as a warm up for the CFL case, and also for its own sake
-- including both the quasi-free response and the collective response
associated with the Goldstone mode.

The interactions of interest are
\begin{equation}
{\mathcal L}_{int} = {\mathcal L}_Z + {\mathcal L}_S.
\label{l_rel}
\end{equation}
Here, ${\mathcal L}_Z$ represents the interactions of neutrinos with
the medium and is given by Eqs.~(\ref{lz2a},\ref{lz2b}).  For
concreteness we use $c_V = -1$ and $c_A = -1.23$.  ${\mathcal L}_S$
represents the strong interactions in the medium that give rise to the
superfluid mode $H$ associated with the breaking of $U(1)_B$.  Since
we expect these interactions to have the form $H(\bar{q} i \gamma_5
q_C + \bar{q_C} i \gamma_5 q)$ (see
Refs. \cite{Casalbuoni:2000na,Rho:1999xf, Rho:2000ww}), we use a
four-quark interaction $(\bar{q} i \gamma_5 q_C + \bar{q_C} i \gamma_5
q )^2.$ To express this in terms of Nambu-Gor'kov spinors, we define
\beq \Gamma_H \equiv \left( \begin{array}{cc} 0 & i \gamma_5 \\ i
\gamma_5 & 0 \end{array} \right).  \eeq
Then the relevant quark self-interactions are
\beq 
\mathcal{L}_S = G \; (\bar{\Psi} \Gamma_H \Psi)^2.
\eeq

\begin{figure}[t]
\centerline{\scalebox{.6}[.6]{\includegraphics{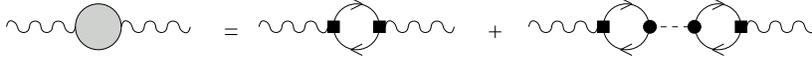}}}
\caption{The two contributions to the medium polarization tensor.  The
first term on the right-hand side of Fig.~\ref{quarkpolarization_Fig} 
corresponds to the quasi-free
response of the medium to the neutrino probe.  The second term on the
right-hand side corresponds to the collective response associated with
the Goldstone mode $H$.}
\label{quarkpolarization_Fig}
\end{figure}

We will calculate two contributions to the medium polarization tensor,
depicted in Fig.~\ref{quarkpolarization_Fig} using the Nambu-Gorkov
propagator given in Eq.~(\ref{NGprop}). Our formalism, notation and 
conventions closely follow those of Ref.~\cite{Rischke:2000qz}. 

\begin{figure}[t]
\centerline{\scalebox{.4}[.4]{\includegraphics{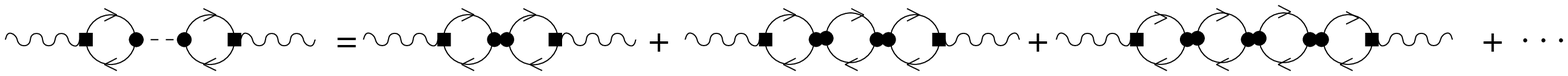}}}
\caption{The collective response of the medium can be evaluated in the
Random Phase Approximation by summing a series of diagrams.}
\label{nu-H_series_Fig}
\end{figure}

The first term on the right-hand side corresponds to the quasi-free
response of the medium.  The second term captures the collective
response of the system associated with the superfluid excitation $H$.
The collective response can be expressed as the RPA sum depicted in
Fig.~\ref{nu-H_series_Fig}.
We will first calculate the quasi-free response, then the collective response.

%\subsection{Quasi-free response}
%
The quasi-free response of the medium comes from first term on the
right-hand side of Fig.~\ref{quarkpolarization_Fig}.  This diagram
makes the following contribution to the polarization tensor:
\begin{eqnarray}\label{rel_qq_result}
\lefteqn{\Pi^{\mu \nu}_{qf} (Q) =} \nonumber\\ & & \frac{1}{2} \int \frac{d^3
p}{(2 \pi)^3} \left[ A(E_{\bf p}, E_{\bf p+q}) \left\{ (c_V^2 + c_A^2)
{\cal T}^{\mu \nu}_+({\bf p}, {\bf p+q}) -2 c_V c_A {\cal W}^{\mu
\nu}_+({\bf p}, {\bf p+q}) \right\} \right.  \nonumber\\ & & \hspace{2cm} +
A(-E_{\bf p}, -E_{\bf p+q}) \left\{ (c_V^2 + c_A^2) {\cal T}^{\mu
\nu}_-({\bf p}, {\bf p+q}) +2 c_V c_A {\cal W}^{\mu \nu}_-({\bf p},
{\bf p+q}) \right\} \nonumber\\ & & \hspace{2cm} - \Delta^2 B(E_{\bf p}, E_{\bf
p+q}) \left\{ (c_V^2 + c_A^2) [{\cal U}^{\mu \nu}_+({\bf p}, {\bf
p+q})+ {\cal U}^{\mu \nu}_-({\bf p}, {\bf p+q})] \right.\nonumber\\ & &
\hspace{5cm} + \left.\left. 2 c_V c_A [{\cal V}^{\mu \nu}_+({\bf p},
{\bf p+q})- {\cal V}^{\mu \nu}_-({\bf p}, {\bf p+q})] \right\}
\right].
\end{eqnarray}
The Matsubara sum led to the quantities
\begin{eqnarray}\label{rel_qq_matsu_1_explicit}
\lefteqn{A(E',E) =} \nonumber\\ 
& & \frac{1}{(2 \xi)(2 \xi')}
\{[n(\xi)-n(\xi')]\left( \frac{(\xi-E)(\xi'-E')}{q_0 + \xi -\xi'} -
\frac{(\xi+E)(\xi'+E')}{q_0 - \xi +\xi'} \right) \nonumber\\ & & \hspace{1.5cm}
+ [1-n(\xi)-n(\xi')]\left( \frac{(\xi+E)(\xi'-E')}{q_0 - \xi -\xi'} -
\frac{(\xi-E)(\xi'+E')}{q_0 + \xi +\xi'} \right)\}
\end{eqnarray} 
and
\begin{eqnarray}\label{rel_qq_matsu_2_explicit}
\lefteqn{B(E',E) =} \nonumber\\ & & \frac{1}{(2 \xi)(2 \xi')}
\{[n(\xi)-n(\xi')]\left( \frac{1}{q_0 + \xi -\xi'} - \frac{1}{q_0 -
\xi +\xi'} \right) \nonumber\\ & & \hspace{1.5cm} +
[1-n(\xi)-n(\xi')]\left( -\frac{1}{q_0 - \xi -\xi'} + \frac{1}{q_0 +
\xi +\xi'} \right)\}
\end{eqnarray}
with
$$ \xi' = \sqrt{{E'}^2 + \Delta^2} \qquad\mbox{and}\qquad \xi =
\sqrt{E^2 + \Delta^2}. $$
The trace over Dirac indices led to the quantities
\begin{eqnarray}\label{rel_qq_T_explicit}
{\cal T}^{\mu \nu}_\pm({\bf p_1}, {\bf p_2}) &=& \left\{
\begin{array}{c@{\quad\mbox{if}\quad}c} 1+ {\bf \widehat{p_1}} \cdot
{\bf \widehat{p_2}} & \mu =0 \;\;\mbox{and}\;\; \nu=0 \\ \pm({\bf
\widehat{p_1} + \widehat{p_2}})^\nu & \mu =0 \;\;\mbox{and}\;\;
\nu\not=0 \\ \pm({\bf \widehat{p_1} + \widehat{p_2}})^\mu & \mu \not=0
\;\;\mbox{and}\;\; \nu=0 \\ \delta^{\mu \nu}(1- {\bf \widehat{p_1}}
\cdot {\bf \widehat{p_2}}) + ({\bf \widehat{p_1}})^\mu ({\bf
\widehat{p_2}})^\nu +({\bf \widehat{p_2}})^\mu ({\bf
\widehat{p_1}})^\nu & \mu \not=0 \;\;\mbox{and}\;\; \nu\not=0
\end{array} \right.
\end{eqnarray}
and
\begin{eqnarray}\label{rel_qq_W_explicit}
{\cal W}^{\mu \nu}_\pm({\bf p_1}, {\bf p_2}) &=& \left\{
\begin{array}{c@{\quad\mbox{if}\quad}c} i \epsilon^{\mu \nu a b} ({\bf
\widehat{p_1}})_a ({\bf \widehat{p_2}})_b & \mu =0 \;\;\mbox{or}\;\;
\nu=0 \\ \mp i \epsilon^{\mu \nu 0 a} ({\bf \widehat{p_1} -
\widehat{p_2}})_a & \mu \not=0 \;\;\mbox{and}\;\;
\nu\not=0
\end{array} \right.
\end{eqnarray}
and
\begin{eqnarray}\label{rel_qq_U_explicit}
{\cal U}^{\mu \nu}_\pm({\bf p_1}, {\bf p_2}) &=& \left\{
\begin{array}{c@{\quad\mbox{if}\quad}c} {\cal T}^{\mu
\nu}_\pm({\bf p_1}, {\bf p_2}) & \nu=0 \\ -
{\cal T}^{\mu \nu}_\pm({\bf p_1}, {\bf p_2}) & \nu\not=0  
\end{array} \right.
\end{eqnarray}
and
\begin{eqnarray}\label{rel_qq_V_explicit}
{\cal V}^{\mu \nu}_\pm({\bf p_1}, {\bf p_2}) &=& \left\{ \begin{array}{c@{\quad\mbox{for}\quad}c} -{\cal W}^{\mu
\nu}_\pm({\bf p_1}, {\bf p_2}) & \nu=0 \\
{\cal W}^{\mu \nu}_\pm({\bf p_1}, {\bf p_2}) & \nu\not=0  
\end{array} \right. .
\end{eqnarray}
Details of this calculation are given in Appendix A.1.

\begin{figure}[t]
\centerline{\scalebox{.6}[.6]{\includegraphics{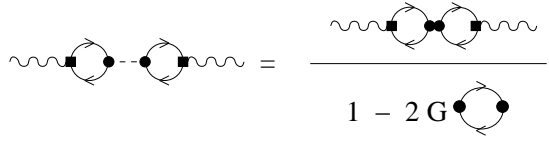}}} 
\caption{The diagrams of the Random Phase Approximation are a
geometric series.}
\label{nu-H_geo_Fig}
\end{figure}

%\subsection{Collective response}
%
Now we want to calculate the contribution to the polarization tensor
from the superfluid mode $H$.  This contribution is the sum of a
series of terms, expressed diagrammatically in
Fig.~\ref{nu-H_series_Fig}.  The series is a geometric series, and its
sum is expressed diagrammatically in Fig.~\ref{nu-H_geo_Fig}.
The contribution of this geometric series to the polarization tensor
is
\begin{equation}
\Pi^{\mu\nu}_H(Q) = -\frac{G}{2} \; \frac{I^\mu(Q) I^\nu(-Q)}{g(Q)},
\end{equation}
where
\begin{eqnarray}
I^\mu(Q) &=& \int\frac{d^3p}{(2 \pi)^3} \; i \Delta \times \nonumber\\
& &\hspace{-.6cm} c_V \left[ C(E_{\bf p}, E_{\bf p+q}) \; {\cal
U}^{\mu 0}_+({\bf p},{\bf p+q}) + C(-E_{\bf p}, -E_{\bf p+q}) \; {\cal
U}^{\mu 0}_-({\bf p},{\bf p+q}) \right]
\label{rel_I_final}
\end{eqnarray}
with
\begin{eqnarray}\label{rel_H_matsu34}
C(E',E) &=& \frac{1}{(2 \xi)(2 \xi')} \{[n(\xi)-n(\xi')]\left(
\frac{E-E'-\xi+\xi'}{q_0 + \xi -\xi'} - \frac{E-E'+\xi-\xi'}{q_0 - \xi
+\xi'} \right) \nonumber\\ & & \hspace{.7cm} +
[1-n(\xi)-n(\xi')]\left( -\frac{E-E'+\xi+\xi'}{q_0 - \xi -\xi'} +
\frac{E - E' - \xi - \xi'}{q_0 + \xi +\xi'} \right)\} ,
\end{eqnarray}
and
\begin{eqnarray}
g(Q)  &=&
1+ 2 \: G \int \frac{d^3p}{(2
\pi)^3}(1+{\bf \hat{p}} \cdot {\bf \widehat{p+q}}) \nonumber\\ & & \times\frac{1}{(2\xi_{\bf
p+q})(2\xi_{\bf p} )}\{\left[n(\xi_{\bf p+q})-n(\xi_{\bf
p})\right]\left(\xi_{\bf p+q} \xi_{\bf p} - E_{\bf p+q} E_{\bf p}
-\Delta^2\right) \nonumber\\ & & \hspace{4.5cm} \left(\frac{1}{q_0+\xi_{\bf
p+q}-\xi_{\bf p}} - \frac{1}{q_0-\xi_{\bf p+q}+\xi_{\bf p}}\right) \nonumber\\
& & \hspace{2.3cm}+\left[1-n(\xi_{\bf p+q})-n(\xi_{\bf
p})\right]\left(\xi_{\bf p+q} \xi_{\bf p} +E_{\bf p+q} E_{\bf p}
+\Delta^2\right) \nonumber\\ & & \hspace{4.5cm} \left(\frac{1}{q_0-\xi_{\bf
p+q}-\xi_{\bf p}} - \frac{1}{q_0+\xi_{\bf p+q}+\xi_{\bf p}}\right) \}. \label{rel_g_explicit}
\end{eqnarray}
\\ Details of the evaluation are given in Appendix A.2.  We should
check that $g(Q)$ vanishes for $q_0=0$ and ${\bf q}=0$; this will
indicate the presence of a massless excitation, namely, the $H$ boson.
Setting $q_0 = q =0$, we need
\beq 
0 = 1 - 2 \: G \int \frac{d^3p}{(2 \pi)^3} \left[1-2 n(\xi_{\bf p})\right] \frac{1}{\xi_{\bf p}} .
\eeq
This easily verified by noting it is the gap equation obtained by minimizing the free energy: $\partial \Omega / \partial \Delta=0$. 

\begin{figure}[t]
\begin{center}
\includegraphics[width=14cm]{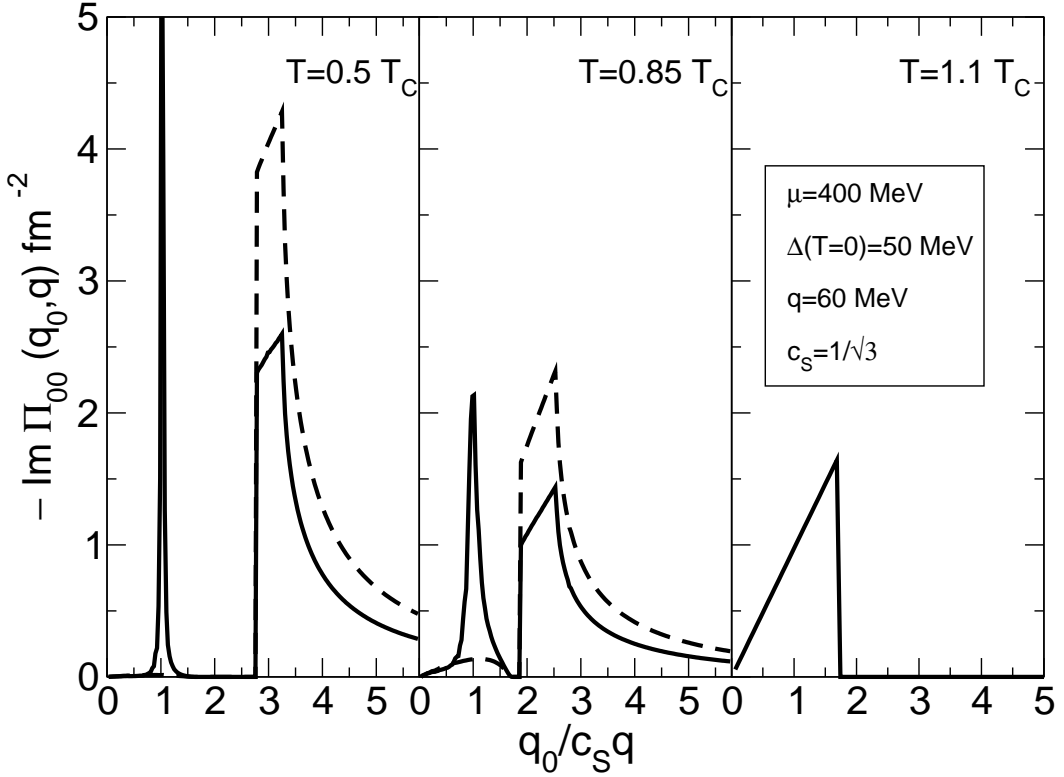}
%\centerline{\epsfxsize 14cm \epsffile{impiz_rel.eps}}
\end{center}
\caption{We plot -Im$\Pi^{00}$ for a one-component, relativistic
superfluid at various temperatures, as a function of the dimensionless
variable $q_0 / (q c_s)$, where $c_s = 1/\sqrt{3}$, with fixed $q =
60$ MeV. The dashed line is the quasi-free response, and the solid
line is the full (that is, quasi-free plus collective) response.  We
use chemical potential $\mu = 400$ MeV.  The gap at zero temperature
is $\Delta = 50$ MeV, and the critical temperature is $T_c = 28.3$
MeV.}
\label{rel_pi00im_Fig}
\end{figure}

%\subsection{Total response}
%
In Fig.~\ref{rel_pi00im_Fig} we plot -Im $\Pi^{00}$ for various
temperatures as a function of $q_0 / (q c_s)$, where $c_s =
1/\sqrt{3}$, which is the expected velocity of the Goldstone mode.
The dashed line is the quasi-free response, and the solid line is the
full (that is, quasi-free plus collective) response.  The momentum
transferred to the medium is fixed at $q=60$ MeV.  We use chemical
potential $\mu = 400$ MeV.  The gap at zero temperature is $\Delta =
50$ MeV, and the critical temperature is $T_c = 28.3$ MeV.  The upper
left panel illustrates $T/T_c = 0.50$.  At this temperature the gap is
$\Delta = 47.9$ MeV.  We find a narrow peak about $q_0 / (q c_s) = 1$,
confirming our expectation about the Goldstone mode.  The threshold
seen at $q_0 / (q c_s) = 2.76$ corresponds to $q_0 = 95.8~\mbox{MeV} =
2 \Delta$. The middle panel illustrates $T/T_c = 0.85$.  At this
temperature the gap is $\Delta = 32.2$ MeV.  The peak associated with 
the Goldstone mode has widened, and
its height has shrunk relative to the response at $q_0 = 2\Delta$.
Also note the nonzero response for $q_0 / (q c_s) \le 1.73$,
corresponding to $ q_0 \leq q$.  The rightmost panel
illustrates $T/T_c = 1.10$.  At this temperature the gap is $\Delta =
0$.  Only the $q_0 \leq q$ response is seen, as expected.
Having studied the medium polarization tensor for a one-component,
relativistic superfluid, we turn to the case of a multi-component,
relativistic superfluid -- color-flavor locked quark matter (CFL).
\section{Response of CFL}
We finally study CFL.  The interactions can still be written as
Eq.~(\ref{l_rel}), except that for three-flavor quark matter we should
use
\beq
c_V = \left( \begin{array}{ccc} \frac{1}{2} - \frac{4}{3}
\sin^2\theta_W & 0 & 0 \\ 0 & -\frac{1}{2}+\frac{2}{3} \sin^2\theta_W
& 0 \\0 & 0 & -\frac{1}{2}+\frac{2}{3} \sin^2\theta_W \end{array}
\right),
\eeq
where the weak mixing angle is $\sin^2\theta_W \approx
0.231$, and
\beq
c_A = \left( \begin{array}{ccc} \frac{1}{2} & 0 & 0 \\ 0 & -\frac{1}{2} & 0 \\0 & 0 & -\frac{1}{2} \end{array} \right).
\eeq
Also, the four-quark interaction that gives rise to the $H$ in CFL is
\beq
{\mathcal L}_S = G  (\bar{\Psi} \Gamma_H \Psi)^2,
\eeq
using
\beq
\Gamma_H = \left( \begin{array}{cc} 0 & i \gamma_5 M \\ i \gamma_5
M & 0 \end{array} \right) 
\qquad \mbox{with} \qquad
(M)^{cd}_{fg} = \epsilon^{cdI} \; \epsilon^{fgI},
\eeq
where $c,d$ are color indices, and $f,g$ are flavor indices.
In CFL the nine quarks form an octet with gap $\Delta_8 \equiv \Delta$
and a singlet with gap $\Delta_1 \equiv 2 \Delta.$  
We are neglecting condensation in
the color ${\bf 6}$ channel, keeping only the condensate in the 
${\bf \bar{3}}$ channel.  The propagator involves the energies
$ \xi_{m \; {\bf p}} = \sqrt{(p-\mu)^2 + \Delta_m^2}.$  
The propagator also involves the following color-flavor matrices:
$$
({\bf P_1})^{cd}_{fg} = ({\bf \tilde{P}_1})^{cd}_{fg} = \frac{1}{3}
\delta^c_f \delta^d_g,
~~~
({\bf P_8})^{cd}_{fg} = \delta_{fg} \delta^{cd} - \frac{1}{3}
\delta^c_f \delta^d_g,
~~~\mbox{and}~~~
({\bf \tilde{P}_8})^{cd}_{fg} = -\delta^d_f \delta^c_g +
\frac{1}{3} \delta^c_f \delta^d_g.
$$
Now we can write down the quark propagator:
$$ S(P) = \left( \begin{array}{cc} G^+(P) & \Xi^-(P) \\ \Xi^+(P) &
G^-(P) \end{array} \right),$$ where
$$ G^+(P) = \left\{ \frac{{\bf P_1}}{p_0^2 - \xi_{1 \; {\bf p}}^2 } +
\frac{{\bf P_8}}{p_0^2 - \xi_{8 \; {\bf p}}^2 } \right\} (p_0 +
E_{\bf p}) \Lambda^+_{\bf p} \gamma_0,$$
$$ G^-(P) = \left\{ \frac{{\bf P_1}}{p_0^2 - \xi_{1 \; {\bf p}}^2 } +
\frac{{\bf P_8}}{p_0^2 - \xi_{8 \; {\bf p}}^2 } \right\} (p_0 -
E_{\bf p}) \Lambda^-_{\bf p} \gamma_0,$$
$$ \Xi^+(P) = - \left\{ \frac{\Delta_1 {\bf \tilde{P}_1} }{p_0^2 - \xi^2_{1 \;
{\bf p}}} + \frac{\Delta_8 {\bf \tilde{P}_8} }{p_0^2 - \xi^2_{8 \;
{\bf p}}} \right\} \gamma_5 \Lambda^-_{\bf p},$$
and
$$ \Xi^-(P) = \left\{ \frac{\Delta_1 {\bf \tilde{P}_1} }{p_0^2 - \xi^2_{1 \;
{\bf p}}} + \frac{\Delta_8 {\bf \tilde{P}_8} }{p_0^2 - \xi^2_{8 \;
{\bf p}}} \right\} \gamma_5 \Lambda^+_{\bf p}.$$ 
We have neglected the contribution from antiquarks.

We now calculate the contributions to the medium polarization tensor
from the quasi-free response and the collective response, the two terms 
depicted on the right-hand side of Fig.~\ref{quarkpolarization_Fig}.

%\subsection{Quasi-free response}
%
The quasi-free contribution to the polarization tensor comes from the
first term on the right-hand side of Fig.~\ref{quarkpolarization_Fig},
and its value is
\begin{eqnarray}
\Pi^{\mu \nu}_{qf}(Q)
& = & \frac{1}{2} \int \frac{d^3 p}{(2 \pi)^3} \sum_{m,n \in
\{1,8\}}\left( \bar{A}(E_{\bf p}, E_{\bf p+q}, \Delta_m,\Delta_n) \left\{
R^{(1)}_{mn} \; {\cal T}^{\mu \nu}_+({\bf p}, {\bf p+q}) -
R^{(2)}_{mn} \; {\cal W}^{\mu \nu}_+({\bf p}, {\bf p+q}) \right\}
\right.  \nonumber \\ 
& & \hspace{3cm} + \bar{A}(-E_{\bf p}, -E_{\bf p+q}, \Delta_m,
\Delta_n) \left\{ R^{(1)}_{mn} \; {\cal T}^{\mu \nu}_-({\bf p}, {\bf
p+q}) + R^{(2)}_{mn} \; {\cal W}^{\mu \nu}_-({\bf p}, {\bf p+q})
\right\} \nonumber \\ 
& & \hspace{3cm} -\Delta_m \Delta_n \bar{B}(E_{\bf p}, E_{\bf
p+q}, \Delta_m, \Delta_n) \left\{ \tilde{R}^{(1)}_{mn} \; [{\cal
U}^{\mu \nu}_+({\bf p}, {\bf p+q}) + {\cal U}^{\mu \nu}_-({\bf p},
{\bf p+q})] \right. \nonumber \\ 
& & \hspace{7.5cm} +
\left.\left. \tilde{R}^{(2)}_{mn} \; [{\cal V}^{\mu \nu}_+({\bf p},
{\bf p+q})- {\cal V}^{\mu \nu}_-({\bf p}, {\bf p+q})] \right\}
\right). \nonumber \\
& & 
\end{eqnarray}
The Matsubara sum led to the quantities $\bar{A}$ and $\bar{B}$, where
\begin{eqnarray}
\lefteqn{\bar{A}(E',E,\Delta',\Delta) =} \nonumber \\ 
& & \frac{1}{(2 \xi)(2 \xi')}
\{[n(\xi)-n(\xi')]\left( \frac{(\xi-E)(\xi'-E')}{q_0 + \xi -\xi'} -
\frac{(\xi+E)(\xi'+E')}{q_0 - \xi +\xi'} \right) \nonumber \\ 
& & \hspace{1.5cm}
+ [1-n(\xi)-n(\xi')]\left( \frac{(\xi+E)(\xi'-E')}{q_0 - \xi -\xi'} -
\frac{(\xi-E)(\xi'+E')}{q_0 + \xi +\xi'} \right)\}
\end{eqnarray} 
and
\begin{eqnarray}
\lefteqn{\hspace{-1.5cm}\bar{B}(E',E,\Delta',\Delta) =} \nonumber \\ 
& & \hspace{-1.5cm}\frac{1}{(2 \xi)(2 \xi')} \{[n(\xi)-n(\xi')]\left(
\frac{1}{q_0 + \xi -\xi'} - \frac{1}{q_0 - \xi +\xi'} \right) \nonumber \\ 
& & + [1-n(\xi)-n(\xi')]\left( -\frac{1}{q_0 - \xi -\xi'} + \frac{1}{q_0 +
\xi +\xi'} \right)\}
\end{eqnarray}
with
$$ \xi' = \sqrt{(E')^2 + (\Delta')^2} \qquad\mbox{and}\qquad \xi =
\sqrt{E^2 + \Delta^2}. $$
The trace over color-flavor indices led to the quantities
\begin{equation*}
R^{(1)}_{mn} = \left( \begin{array}{cc} 0.0556 & 0.2865 \\0.2865  & 2.4502
 \end{array} \right)\,,
\quad
R^{(2)}_{mn} = \left( \begin{array}{cc} 0.0556 & 0.2391 \\0.2391  & 2.1182
 \end{array} \right)\,,
\end{equation*}
\begin{equation}
\tilde{R}^{(1)}_{mn} = \left( \begin{array}{cc} 0.0556 & -0.2865 \\ -0.2865 &
 -0.1286\end{array} \right)\,,
\; \;
\mbox{and}
\; \; \;
\tilde{R}^{(2)}_{mn} = \left( \begin{array}{cc} 0.0556 & -0.2391 \\ -0.2391 &-0.0338
 \end{array} \right)\,.
\label{cf_free}
\end{equation}
Note that these matrices include off-diagonal components: the weak
interactions can scatter the singlet fermionic quasi-particle into one
of the octet fermionic quasi-particles, and vice-versa.  We discuss
the color-flavor trace further in Appendix B.1.

%\subsection{Collective response}
We now want to calculate the contribution to the quark polarization
tensor from the collective response of the medium.  This is depicted
in Fig.~\ref{nu-H_geo_Fig}.  Its value is
\begin{equation}
\Pi^{\mu\nu}_H(Q) = -\frac{G}{2} \; \frac{\bar{I}^\mu(Q) \bar{I}^\nu(-Q)}{\bar{g}(Q)},
\end{equation}
where
\begin{eqnarray}
\bar{I}^\mu(Q) & = & \int \frac{d^3 p}{(2 \pi)^3} \sum_{m \in \{1,8\}} i
\Delta_m \; R^{(3)}_m  \left[ \bar{C}(E_{\bf p}, E_{\bf p+q},
\xi_{m \; {\bf p}},\xi_{m \; {\bf p+q}}) \; {\cal U}^{\mu
0}_+({\bf p}, {\bf p+q}) \right.  \nonumber \\ 
& & \hspace{4.25cm} + \left. \bar{C}(-E_{\bf p}, -E_{\bf p+q}, \xi_{m \; {\bf p}},\xi_{m
\; {\bf p+q}})\; {\cal U}^{\mu 0}_-({\bf p}, {\bf p+q}) \right],
\label{cfl_I_explicit}
\end{eqnarray}
with
\begin{eqnarray}
\lefteqn{\bar{C}(E',E,\xi',\xi) = } \nonumber \\ 
& & \frac{1}{(2 \xi)(2 \xi')} \{[n(\xi)-n(\xi')]\left( \frac{-E+E'+\xi-\xi'}{q_0 + \xi
-\xi'} - \frac{-E+E'-\xi+\xi'}{q_0 - \xi +\xi'} \right) \nonumber \\ 
& & \hspace{1.5cm} + [1-n(\xi)-n(\xi')]\left( -\frac{-E+E'-\xi-\xi'}{q_0 -
\xi -\xi'} + \frac{-E+E'+\xi+\xi'}{q_0 + \xi +\xi'} \right)\}.
\end{eqnarray}
The quantity $R^{(3)}_m$ arises from the color-flavor trace.  Its value is
\begin{equation}
R^{(3)}_1 = -\frac{1}{3} \qquad \mbox{and} \qquad R^{(3)}_8 =
-\frac{4}{3}.
\label{cf_I}
\end{equation}
Also,
\begin{equation}
\bar{g}(Q) = 1 + 2 \; G \int \frac{d^3 p}{(2 \pi)^3} (1 + {\bf \hat{p}} \cdot
{\bf \widehat{p+q}})\sum_{m} \bar{D}(E_{\bf p}, E_{\bf p+q}, \Delta_m) \;
R^{(5)}_m,
\label{cfl_g_explicit}
\end{equation}
where
\begin{eqnarray}
\lefteqn{\bar{D}(E',E,\Delta) = } \nonumber \\ 
& & \frac{1}{(2 \xi)(2 \xi')}
\left\{ [n(\xi)-n(\xi')] (\xi \xi' - E E' - \Delta^2) \left(
\frac{1}{q_0 + \xi -\xi'} - \frac{1}{q_0 - \xi +\xi'} \right)
\right. \nonumber \\ 
& & \hspace{1cm} + \left. [1-n(\xi)-n(\xi')] (\xi \xi' +
E E' + \Delta^2) \left( \frac{1}{q_0 - \xi -\xi'} - \frac{1}{q_0 +
\xi +\xi'} \right) \right\}.
\end{eqnarray}
and
\begin{equation}
R^{(5)}_1 = 4 \qquad \mbox{and} \qquad R^{(5)}_8 = 8.
\label{cf_g}
\end{equation}
We discuss the color-flavor traces further in Appendix B.2.

The quantity $g(Q)$ should vanish when $q_0=q=0$ in order to 
describe a massless excitation.  That is, we must have
\beq
0 = 1 - 2 \; G \int \frac{d^3p}{(2 \pi)^3} \left\{[1-2n(\xi_{1 \; {\bf
p}})]\frac{4}{\xi_{1 \; {\bf p}}} + [1-2n(\xi_{8 \; {\bf
p}})]\frac{8}{\xi_{8 \; {\bf p}}} \right\}.
\eeq
This is the gap equation.  It gives a critical temperature $T_c
\approx 0.71 \; \Delta(T=0)$, in agreement with
Ref.~\cite{Schmitt:2002sc}.

%\subsection{Total response and differential cross section}
\begin{figure}[h]
\begin{center}
\includegraphics[width=14cm]{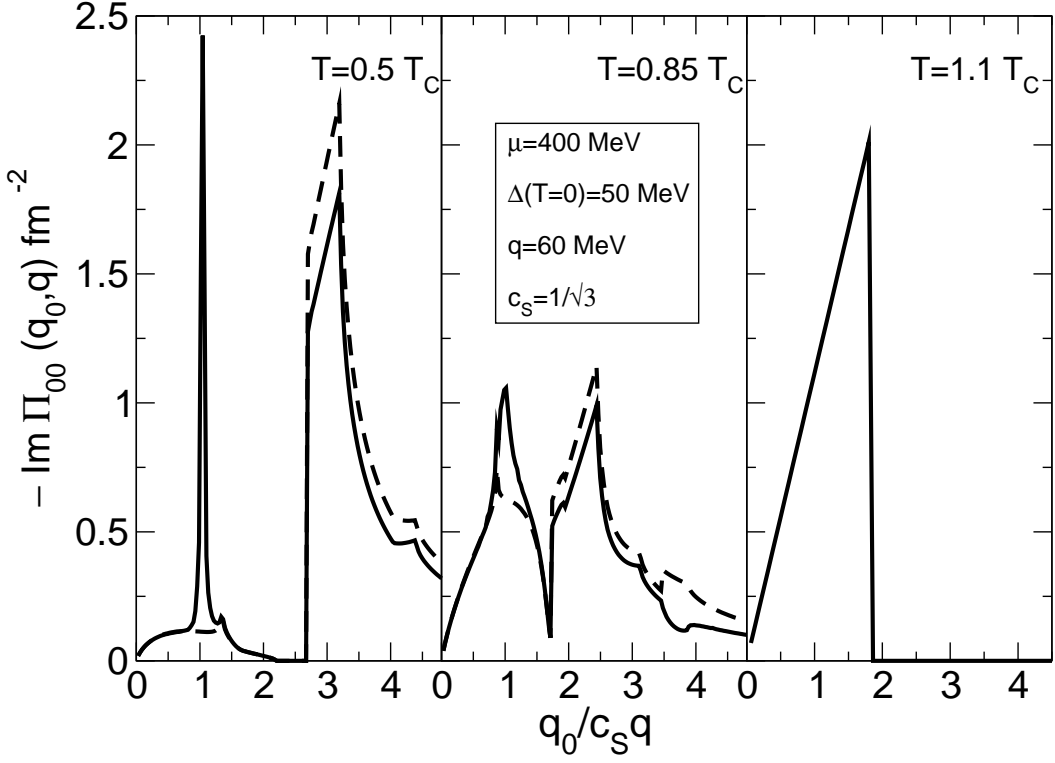}
%\centerline{\epsfxsize 14cm \epsffile{impiz_cfl.eps}}
\end{center}
\caption{We plot -Im$\Pi^{00}$ for CFL at various temperatures as a
function of $q_0 / (q c_s)$, where $c_s = 1/\sqrt{3}$, with fixed $q =
60$ MeV. The dashed line is the quasi-free response, and the solid
line is the full (that is, quasi-free plus collective) response.  We use
quark chemical potential $\mu = 400$ MeV.  The gap at zero temperature
is $\Delta = 50$ MeV, and the critical temperature is $T_c = 35.7$
MeV.}
\label{cfl_pi00im_Fig}
\end{figure}
In Fig.~\ref{cfl_pi00im_Fig} we plot -Im $\Pi^{00}$ for various
temperatures as a function of $q_0 / (q c_s)$, where $c_s =
1/\sqrt{3}$, which is the expected velocity of the Goldstone mode.
The dashed line is the quasi-free response, and the solid line is the
full (that is, quasi-free plus collective) response.  The momentum
transferred to the medium is fixed at $q=60$ MeV.  We use quark
chemical potential $\mu = 400$ MeV.  The gap at zero temperature is
$\Delta = 50$ MeV, and the critical temperature is $T_c = 35.7$ MeV.
The leftmost panel illustrates $T/T_c = 0.50$.  At this temperature
the gap is $\Delta = 46.7$ MeV.  As in the one-component case, we find
a narrow peak about $q_0 / (q c_s) = 1$, confirming our expectation
about the Goldstone mode.  There is a threshold at $q_0 / (q c_s) =
2.69$, corresponding to $q_0 = 2 \Delta$, and there is an additional
small response for $q_0/(q c_s) \le 1.73$, that is, $q_0 \le q$.  The
CFL response function is much more busy than in the one-component case
(see Fig.~\ref{rel_pi00im_Fig}) because of additional effects at $q_0
= \Delta, 3\Delta, 4\Delta$, arising from the fact that some fermionic quasi-particles  in
CFL matter have a gap $\Delta$ and some have a gap $2\Delta$.  The upper
right panel illustrates $T/T_c = 0.85$.  At this temperature the gap
is $\Delta = 30.0$ MeV.  The peak associated with the Goldstone mode has widened, and its height has
shrunk relative to the response at $q_0 = 2\Delta$, but it overlaps
with the $q_0 \le q$ response, which has grown.  The lower panel illustrates $T/T_c = 1.10$.  At this temperature the gap
is $\Delta = 0$.  Only the $q_0 \leq q$ response is seen, as expected.
\begin{figure}[h]
\begin{center}
\includegraphics[width=14cm]{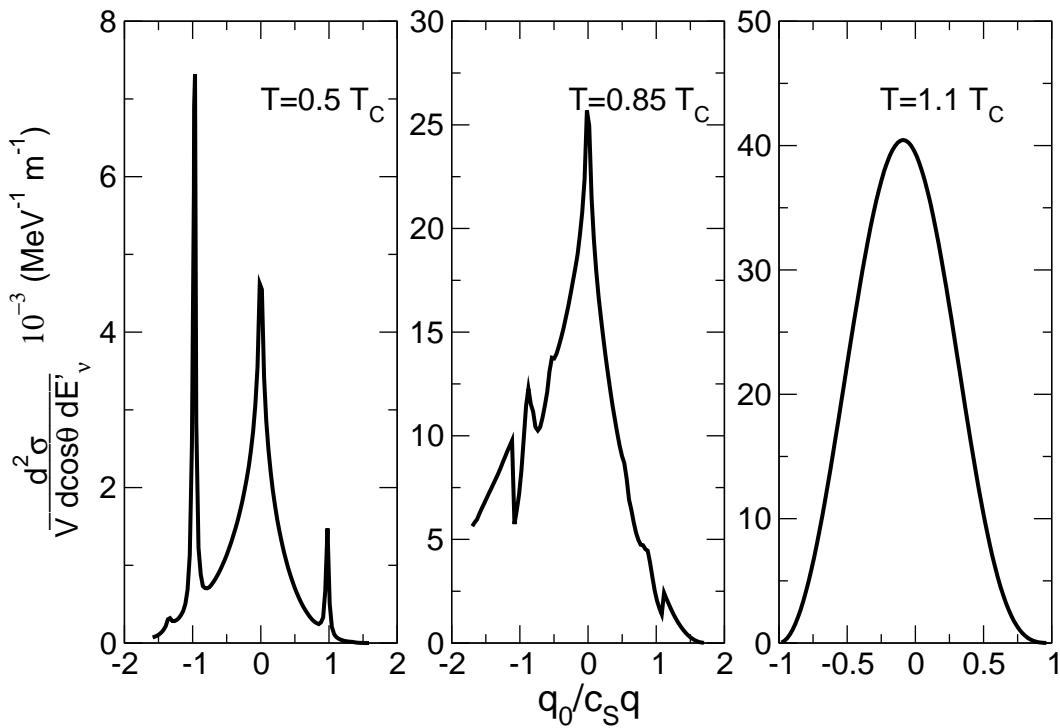}
%\centerline{\epsfxsize 14cm \epsffile{dsigma_cfl.eps}}
\end{center}
\caption{We plot the differential cross section for neutral-current neutrino
scattering in CFL as a function of the dimensionless variable $q_0/(q
c_s)$, with $c_s = 1/\sqrt{3}$ and with momentum transfer set to the typical incoming neutrino energy $q = E_\nu = \pi T$.  
The spike at $q_0/(q c_s)\simeq1$ is the Goldstone mode
response. We use quark chemical potential $\mu = 400$
MeV.  The gap at zero temperature is $\Delta = 50$ MeV, and the critical temperature is $T_c = 35.7$ MeV. 
}
\label{cfl_dsigma_Fig}
\end{figure}

In Fig.~\ref{cfl_dsigma_Fig} we plot the differential cross section
for neutral-current neutrino scattering in CFL.  The energy of the
incoming neutrino is $E_\nu = \pi T$, the typical energy of a thermal
neutrino.  The momentum transferred to the medium is set to $q =
E_\nu$.  Kinematics demand $-1.73 \leq q_0/(q c_s) \leq 1.73$.  The
parameters are otherwise as in Fig.~\ref{cfl_pi00im_Fig}.  The
leftmost panel is again $T/T_c = 0.50$.  There are prominent peaks
near $q_0/(q c_s) = \pm 1$ corresponding to the Goldstone mode.  The
expected feature at $q_0 \geq 2\Delta$, corresponding to excitation of
a particle-hole pair, does not lie in the kinematically allowed region
of $q_0$.  There is a small peak, however, at $q_0/(q c_s) = -1.44$,
or $-q_0 = \Delta$, corresponding to scattering of a singlet quark,
whose gap is $2\Delta$, into an octet quark, whose gap is $\Delta$.
The middle panel illustrates $T/T_c = 0.85$.  The Goldstone peak has
widened, and the threshold at $q_0 \geq 2\Delta$ -- corresponding to
$q_0/(q c_s) \geq 1.09$ -- now falls in the kinematically allowed
region.  The rightmost panel illustrates $T/T_c = 1.10$.  This is the
result for free quarks.
\section{Conclusion}
We have calculated the differential cross section for neutral-current
neutrino scattering in superfluid neutron matter, plotted in
Fig.~\ref{nrdiffcs} and in color-flavor locked quark matter, plotted
in Fig.~\ref{cfl_dsigma_Fig}, under conditions relevant to
proto-neutron stars.  Our results apply above and below the critical
temperature.  In both of these regimes our model for the interaction
in the medium includes the dominant contribution to the cross section.
Above the critical temperature, this comes from the fermionic
excitations, which become the pair-breaking excitations below the
critical temperature.  Well below the critical temperature the
dominant contribution comes from the massless bosonic mode associated
with the breaking of $U(1)$.  Although we presented results results
for scattering cross sections with space-like kinematics, our
polarization functions extend into the time-like region where
pair-breaking (and recombination) is the dominant source of the
response. These could be employed in calculations of the neutrino
emissivity. In particular, we have demonstrated the importance of
vertex corrections in these region.  Our results suggest that earlier
calculations of the neutrino emissivity from the pair recominbation
process in superfluid neutron matter \cite{Flowers:1976} and quark matter
\cite{Jaikumar:2001hq}, which ignore these vertex (RPA)
corrections, need to be revised.

The analysis presented here is based on mean-field and the random
phase approximation (RPA). Its validity is restricted to weak
coupling, $\Delta \ll \mu$. For strong coupling we can expect the
response to differ quantitatively. In particular, screening
corrections which were discussed earlier could be significant.
Another drawback is our use of simplified interactions to describe
superfluid neutron and quark matter. Our focus was to explore the role
of superfluidity and this motivated our choice of a simple zero-range
s-wave interaction.  In reality, the nucleon-nucleon and quark-quark
interactions are more complex.  These will induce
additional correlations which will affect both the gap equation and
the response.  (For a recent review on the role of strong interaction
correlations on neutrino opacities see Ref.~\cite{Burrows:2004vq}.)
Given the non-perturbative nature of these corrections it is difficult
to foresee how large they may be.

Although our results are valid both above and below the critical
temperature, we have implicitly assumed that the transition is a
BCS-like second order transition. Several caveats must be borne in
mind when using our results near $T_c$. In the real system this
transition may be first order either due to gauge field fluctuations
\cite{Giannakis:2001wz} or stresses such as the strange quark mass and
electric charge neutrality \cite{Steiner:2002gx}.  Also, fluctuations
of the magnitude of the order parameter dominate in a region called
the Ginsburg region around $T_c$. In strong coupling, the size of this
region could be significant fraction of $T_c$
\cite{Voskresensky:2003wd}.  Our approach captures some of these
fluctuations through RPA. Nonetheless, a Landau-Ginsburg approach - an
effective theory for $|\Delta|$, is more appropriate in this regime
\cite{Iida:2000ha}. In particular, there are precursor fluctuations
just above $T_c$, which are not included in our response, that may be
relevant \cite{Kitazawa:2001ft}. These effects are currently under
investigation and will be reported elsewhere.

Our primary goal was to provide expressions for the differential cross sections  that could be used in
simulations of the early thermal evolution of neutron stars born in
the aftermath of a supernova explosion.  The microphysics of neutrino
scattering affects the rate of diffusion, which in turn affects
macroscopic observables such as the cooling rate and the neutrino
emission from core-collapse supernovae.  Our results, which extends
both to the low and high temperature regions, are well suited
for use in simulations of core-collapse supernovae and early thermal
evolution of neutron stars.

%  Acknowledgments
%%%%%%%%%%%%%%%%%%
\vskip0.5in \centerline{\bf Acknowledgments} We would like to thank
B. Fore, M. Forbes, K. Fukushima, C. Kouvaris, K. Rajagopal and
G. Rupak for helpful discussions.  J.K. would also like to thank the
Institute for Fundamental Theory at the University of Florida for its
hospitality.  The research of J.K. is supported by the Department of
Energy under cooperative research agreement DE-FC02-94ER40818.  The
research of S. R is supported by the Department of Energy under
contract W-7405-ENG-36.

\appendix{}
\section{Calculation of the polarization tensor for the relativistic superfluid}
In this appendix we detail the calculation of the medium polarization
tensor for the one-component, relativistic superfluid.  In Section A.1
we evaluate the contribution from the quasi-free response function.
In Section A.2 we include the contribution from the collective
response.

\subsection{Quasi-free response}

The contribution to the medium polarization tensor can be associated
with the first diagram on the right-hand side of
Fig.~\ref{quarkpolarization_Fig}.  That diagram has the value
\begin{equation}\label{rel_qq_first}
 \Pi^{\mu \nu}_{qf} (Q) = -i\frac{1}{2}\int \frac{d^4 P}{(2 \pi)^4}
\mbox{Tr}[\Gamma^\mu_Z S(P) \Gamma^\nu_Z S(P+Q)].
\end{equation} 
The factor of $\frac{1}{2}$ is due to the doubling of fermion degrees
of freedom in the Nambu-Gor'kov formalism.  Evaluating the trace over
Nambu-Gor'kov indices gives
\begin{eqnarray}\label{rel_qq_NG}
\Pi^{\mu \nu}_{qf} (Q) &=& -i\frac{1}{2} \int \frac{d^4 P}{(2 \pi)^4}
\mbox{Tr} [ \gamma^\mu (c_V-c_A \gamma_5) G^+(P) \gamma^\nu(c_V-c_A
\gamma_5) G^+(P+Q) \nonumber\\ & & \hspace{2.2cm} -\gamma^\mu (c_V-c_A
\gamma_5) \Xi^-(P) \gamma^\nu(c_V+c_A \gamma_5) \Xi^+(P+Q) \nonumber\\ & &
\hspace{2.2cm} -\gamma^\mu (c_V+c_A \gamma_5) \Xi^+(P)
\gamma^\nu(c_V-c_A \gamma_5) \Xi^-(P+Q) \nonumber\\ & & \hspace{2.2cm}
+\gamma^\mu (c_V+c_A \gamma_5) G^-(P) \gamma^\nu(c_V+c_A \gamma_5)
G^-(P+Q) ],
\end{eqnarray}
where the remaining trace is over Dirac indices. We will write down
the value of each of the four traces above.  First,
\begin{eqnarray*}
\lefteqn{\mbox{Tr}[\gamma^\mu (c_V-c_A \gamma_5) G^+(P)
\gamma^\nu(c_V-c_A \gamma_5) G^+(P+Q)] } \\ 
&=& \frac{(p_0 +
E_{\bf p})(p_0 +q_0+ E_{\bf p+q})}{[p_0^2-\xi_{\bf
p}^2][(p+q)_0^2-\xi_{\bf p+q}^2]} \left\{ (c_V^2 + c_A^2) {\cal
T}^{\mu \nu}_+({\bf p}, {\bf p+q}) -2 c_V c_A {\cal W}^{\mu
\nu}_+({\bf p}, {\bf p+q}) \right\},
\end{eqnarray*}
\\
where ${\cal T}^{\mu \nu}_\pm$ is defined by
\begin{equation}\label{rel_qq_T_def}
{\cal T}^{\mu \nu}_\pm({\bf p_1}, {\bf p_2}) = \mbox{Tr}[\gamma_0
\gamma^\mu \Lambda^{\pm}_{\bf p_1} \gamma_0 \gamma^\nu
\Lambda^{\pm}_{\bf p_2}], 
\end{equation}
and ${\cal W}^{\mu \nu}_\pm$ is defined by
\begin{equation}\label{rel_qq_W_def}
{\cal W}^{\mu \nu}_\pm({\bf p_1}, {\bf p_2}) = \mbox{Tr}[\gamma_0
\gamma^\mu \Lambda^{\pm}_{\bf p_1} \gamma_0 \gamma^\nu
\Lambda^{\pm}_{\bf p_2} \gamma_5].
\end{equation}
Explicit evaluation of ${\cal T}^{\mu \nu}_\pm$ and ${\cal W}^{\mu
\nu}_\pm$ gives Eqs.~(\ref{rel_qq_T_explicit},
\ref{rel_qq_W_explicit}), respectively.  The second term in
(\ref{rel_qq_NG}) is
\begin{eqnarray*}
\lefteqn{\mbox{Tr}[-\gamma^\mu (c_V-c_A \gamma_5) \Xi^-(P)
\gamma^\nu(c_V+c_A \gamma_5) \Xi^+(P+Q)] } \\ 
&=&\frac{-\Delta^2}{[p_0^2-\xi_{\bf p}^2][(p+q)_0^2-\xi_{\bf
p+q}^2]}\left\{ (c_V^2 + c_A^2) {\cal U}^{\mu \nu}_+({\bf p}, {\bf
p+q}) +2 c_V c_A {\cal V}^{\mu \nu}_+({\bf p}, {\bf p+q}) \right\},
\end{eqnarray*}
\\
where ${\cal U}^{\mu \nu}_\pm$ is defined by
\begin{equation}\label{rel_qq_U_def}
{\cal U}^{\mu \nu}_\pm({\bf p_1}, {\bf p_2}) = \mbox{Tr}[\gamma^\mu
\Lambda^{\pm}_{\bf p_1} \gamma^\nu \Lambda^{\mp}_{\bf p_2}],
\end{equation}
and ${\cal V}^{\mu \nu}_\pm$ is defined by
\begin{equation}\label{rel_qq_V_def}
{\cal V}^{\mu \nu}_\pm({\bf p_1}, {\bf p_2}) = \mbox{Tr}[\gamma_0
\gamma^\mu \Lambda^{\pm}_{\bf p_1} \gamma_0 \gamma^\nu
\Lambda^{\pm}_{\bf p_2} \gamma_5].
\end{equation}
Explicit evaulation of ${\cal U}^{\mu \nu}_\pm$ and ${\cal V}^{\mu \nu}_\pm$ gives Eqs.~(\ref{rel_qq_U_explicit}, \ref{rel_qq_V_explicit}), respectively.  The third term in (\ref{rel_qq_NG}) is
\begin{eqnarray*}
\lefteqn{\mbox{Tr}[-\gamma^\mu (c_V+c_A \gamma_5) \Xi^+(P)
\gamma^\nu(c_V-c_A \gamma_5) \Xi^-(P+Q)] } \\ 
&=&\frac{-\Delta^2}{[p_0^2-\xi_{\bf p}^2][(p+q)_0^2-\xi_{\bf
p+q}^2]}\left\{ (c_V^2 + c_A^2) {\cal U}^{\mu \nu}_-({\bf p}, {\bf
p+q}) -2 c_V c_A {\cal V}^{\mu \nu}_-({\bf p}, {\bf p+q}) \right\}.
\end{eqnarray*}
\\
And the final term in (\ref{rel_qq_NG}) is
\begin{eqnarray*}
\lefteqn{\mbox{Tr}[\gamma^\mu (c_V+c_A \gamma_5) G^-(P)
\gamma^\nu(c_V+c_A \gamma_5) G^-(P+Q)] } \\ 
&=& \frac{(p_0 -
E_{\bf p})(p_0 +q_0- E_{\bf p+q})}{[p_0^2-\xi_{\bf
p}^2][(p+q)_0^2-\xi_{\bf p+q}^2]}\left\{ (c_V^2 + c_A^2) {\cal T}^{\mu
\nu}_-({\bf p}, {\bf p+q}) +2 c_V c_A {\cal W}^{\mu \nu}_-({\bf p},
{\bf p+q}) \right\}.
\end{eqnarray*}
\\
Next, we consider the Matsubara sums:
\begin{equation}\label{rel_qq_matsu_1_def}
A(E',E) = \frac{1}{\beta}\sum_n
\frac{(p_0+q_0+E)(p_0+E')}{[(p_0+q_0)^2-\xi^2][p_0^2-\xi'^2]}
\end{equation}
and
\begin{equation}\label{rel_qq_matsu_2_def}
B(E',E) = \frac{1}{\beta}\sum_n \frac{1}{[(p_0+q_0)^2-\xi^2][p_0^2-\xi'^2]},
\end{equation}
where $p_0 = -i (2n+1)\pi/\beta$ is a fermionic Matsubara frequency;
$q_0 = -i 2m \pi/\beta$ is a bosonic Matsubara frequency; and
$$ \xi' = \sqrt{{E'}^2 + \Delta^2} \qquad\mbox{and}\qquad \xi =
\sqrt{E^2 + \Delta^2}. $$ Explicit expressions for $A$ and $B$ are
given in Eqs.~(\ref{rel_qq_matsu_1_explicit},
\ref{rel_qq_matsu_2_explicit}), respectively.  Adding all the terms
yields Eq.~(\ref{rel_qq_result}).

\subsection{Collective response}
In this section we detail the evaluation of the quantity depicted on
the right-hand side of Fig.~\ref{nu-H_geo_Fig}, corresponding to the
collective response of the medium.  First we evaluate the numerator of
that quantity, then the denominator.

\subsubsection{The numerator}
The numerator of the right-hand side of Fig.~\ref{nu-H_geo_Fig} contributes
\begin{eqnarray*}
\lefteqn{- G\: 2\: {1 \over 2} \:{ 1 \over 2}\int
\frac{d^4P}{(2 \pi)^4} \frac{d^4K}{(2 \pi)^4}\mbox{Tr}[\Gamma^\mu_Z
S(P) \Gamma_H S(P+Q)] \; \mbox{Tr}[\Gamma_H S(K) \Gamma^\nu_Z S(K+Q)] = } \hspace{12cm} \\
& &\hspace{-8cm}-\frac{G}{2} I^\mu(Q) I^\nu(-Q),
\end{eqnarray*}
where
\begin{eqnarray}
I^\mu(Q) &\equiv& \int \frac{d^4P}{(2 \pi)^4}\mbox{Tr}[\Gamma^\mu_Z
S(P) \Gamma_H S(P+Q)]. 
\label{rel_I_def}
\end{eqnarray}
To evaluate $I^\mu(Q)$ we first compute
\begin{eqnarray*}
\lefteqn{\mbox{Tr}[\Gamma^\mu_Z S(P) \Gamma_H S(P+Q)] } \\ &=&
\mbox{Tr}[\left( \begin{array}{cc} \gamma^\mu (c_V - c_A \gamma_5) & 0
\\ 0 & -\gamma^\mu (c_V + c_A \gamma_5) \end{array} \right) \left(
\begin{array}{cc} G^+(P) & \Xi^-(P) \\ \Xi^+(P) & G^-(P) \end{array}
\right) \\ & & \hspace{.5cm} \left( \begin{array}{cc} 0 & i \gamma_5
\\ i \gamma_5 & 0 \end{array} \right) \left( \begin{array}{cc}
G^+(P+Q) & \Xi^-(P+Q) \\ \Xi^+(P+Q) & G^-(P+Q) \end{array} \right)] \\
&=&\mbox{Tr}[i \gamma^\mu (c_V - c_A \gamma_5) G^+(P) \gamma_5
\Xi^+(P+Q) \\ & & \hspace{.5cm}+ i \gamma^\mu (c_V - c_A \gamma_5)
\Xi^-(P) \gamma_5 G^+(P+Q) \\ & &\hspace{.5cm} -i \gamma^\mu (c_V +
c_A \gamma_5) \Xi^+(P) \gamma_5 G^-(P+Q) \\ & &\hspace{.5cm} -i
\gamma^\mu (c_V + c_A \gamma_5) G^-(P) \gamma_5 \Xi^-(P+Q) ].
\end{eqnarray*}
\\
Now,
\begin{eqnarray*}
\lefteqn{\mbox{Tr}[i \gamma^\mu (c_V - c_A \gamma_5) G^+(P) \gamma_5
\Xi^+(P+Q)]} \\ &=& \frac{(p_0+E_{\bf p})(-i \Delta)}{[p_0^2-\xi_{\bf
p}^2][(p+q)_0^2-\xi_{\bf p+q}^2]} \left[ c_V {\cal U}^{\mu 0}_+({\bf
p},{\bf p+q}) + c_A {\cal V}^{\mu 0}_+({\bf p},{\bf p+q})\right].
\end{eqnarray*}
Next,
\begin{eqnarray*}
\lefteqn{ \mbox{Tr}[i \gamma^\mu (c_V - c_A \gamma_5) \Xi^-(P)
\gamma_5 G^+(P+Q)]}\\ &=& \frac{(i \Delta)((p+q)_0+E_{\bf
p+q})}{[p_0^2-\xi_{\bf p}^2][(p+q)_0^2-\xi_{\bf p+q}^2]} \left[ c_V
{\cal U}^{\mu 0}_+({\bf p},{\bf p+q}) + c_A {\cal V}^{\mu 0}_+({\bf
p},{\bf p+q})\right].
\end{eqnarray*}
And
\begin{eqnarray*}
\lefteqn{ \mbox{Tr}[-i \gamma^\mu (c_V + c_A \gamma_5) \Xi^+(P)
\gamma_5 G^-(P+Q)]}\\ &=& \frac{(i \Delta)((p+q)_0-E_{\bf
p+q})}{[p_0^2-\xi_{\bf p}^2][(p+q)_0^2-\xi_{\bf p+q}^2]} \left[ c_V
{\cal U}^{\mu 0}_-({\bf p},{\bf p+q}) - c_A {\cal V}^{\mu 0}_-({\bf
p},{\bf p+q})\right] .
\end{eqnarray*}
Finally,
\begin{eqnarray*}
\lefteqn{ \mbox{Tr}[-i \gamma^\mu (c_V + c_A \gamma_5) G^-(P) \gamma_5
\Xi^-(P+Q)]}\\ &=& \frac{(p_0-E_{\bf p})(-i \Delta)}{[p_0^2-\xi_{\bf
p}^2][(p+q)_0^2-\xi_{\bf p+q}^2]} \left[ c_V {\cal U}^{\mu 0}_-({\bf
p},{\bf p+q}) - c_A {\cal V}^{\mu 0}_-({\bf p},{\bf p+q})\right] .
\end{eqnarray*}
\\
To proceed we evaluate the Matsubara sums
\begin{eqnarray*}\label{rel_H_matsu_3}
C_1(E',E) 
&\equiv& \frac{1}{\beta}\sum_n \frac{p_0+E'}{[(p_0+q_0)^2-\xi^2][p_0^2-\xi'^2]} \\
&=& \frac{1}{(2 \xi)(2 \xi')}
\{[n(\xi)-n(\xi')]\left( -\frac{\xi' - E'}{q_0 + \xi -\xi'} -
\frac{\xi'+E'}{q_0 - \xi +\xi'} \right) \nonumber\\ & & \hspace{1.5cm}
+ [1-n(\xi)-n(\xi')]\left( \frac{\xi'-E'}{q_0 - \xi -\xi'} +
\frac{\xi'+E'}{q_0 + \xi +\xi'} \right)\}
\end{eqnarray*}
and
\begin{eqnarray*}\label{rel_H_matsu_4}
C_2(E',E) &\equiv& \frac{1}{\beta}\sum_n \frac{p_0+q_0+E}{[(p_0+q_0)^2-\xi^2][p_0^2-\xi'^2]} \\
&=& \frac{1}{(2 \xi)(2 \xi')}
\{[n(\xi)-n(\xi')]\left( -\frac{\xi - E}{q_0 + \xi -\xi'} -
\frac{\xi+E}{q_0 - \xi +\xi'} \right) \nonumber\\ & & \hspace{1.5cm}
+ [1-n(\xi)-n(\xi')]\left( -\frac{\xi+E}{q_0 - \xi -\xi'} -
\frac{\xi-E}{q_0 + \xi +\xi'} \right)\}
\end{eqnarray*}
\\
If we define
\begin{eqnarray*}
C(E',E) &\equiv& -C_1(E',E)+C_2(E',E) \\
&=& \frac{1}{(2 \xi)(2 \xi')}
\{[n(\xi)-n(\xi')]\left( \frac{E-E'-\xi+\xi'}{q_0 + \xi -\xi'} -
\frac{E-E'+\xi-\xi'}{q_0 - \xi +\xi'} \right) \nonumber\\ & & \hspace{.7cm}
+ [1-n(\xi)-n(\xi')]\left( -\frac{E-E'+\xi+\xi'}{q_0 - \xi -\xi'} +
\frac{E - E' - \xi - \xi'}{q_0 + \xi +\xi'} \right)\}
,
\end{eqnarray*}
then an expression for $I^\mu$ is 
\begin{eqnarray*}
I^\mu(Q)
&=& \int\frac{d^3p}{(2 \pi)^3}  \; i \Delta \times \nonumber\\ &
&\hspace{-.6cm}\left\{ C(E_{\bf p}, E_{\bf p+q}) [ c_V
{\cal U}^{\mu 0}_+({\bf p},{\bf p+q})+ c_A {\cal V}^{\mu 0}_+({\bf
p},{\bf p+q}) ] \right.\nonumber\\ & & \hspace{-.6cm} + \left. C(-E_{\bf p}, -E_{\bf p+q}) [ c_V {\cal U}^{\mu 0}_-({\bf
p},{\bf p+q}) - c_A {\cal V}^{\mu 0}_-({\bf p},{\bf p+q}) ]
\right\}.
\end{eqnarray*}
After performing the $\int d^3p$ integral, the terms proportional to $c_A$ vanish, leaving
\begin{eqnarray*}
I^\mu(Q)
&=& \int\frac{d^3p}{(2 \pi)^3}  \; i \Delta \; c_V \left[ C(E_{\bf p}, E_{\bf p+q}) \; {\cal U}^{\mu 0}_+({\bf p},{\bf p+q}) + C(-E_{\bf p}, -E_{\bf p+q}) \; {\cal U}^{\mu 0}_-({\bf
p},{\bf p+q}) \right].
\end{eqnarray*}
This is Eq.~(\ref{rel_I_final}).

\subsubsection{The denominator}

In this section we evaluate the denominator of the right-hand side of
Fig.~\ref{nu-H_geo_Fig}.  It has the value
\begin{eqnarray*}
g(Q) &\equiv& 1 + G \: 2 \: \frac{1}{2} \int \frac{d^4P}{(2
\pi)^4} \mbox{Tr}[S(P+Q) \Gamma_H S(P) \Gamma_H]  \label{rel_g_def}\\
&=& 1 + G \: 2 \: \frac{1}{2} \int \frac{d^4P}{(2
\pi)^4} \mbox{Tr}[S(P+Q) \Gamma_H S(P) \Gamma_H]  \\ &=& 1+
G \int \frac{d^4P}{(2 \pi)^4} \mbox{Tr}[\left(
\begin{array}{cc} G^+(P+Q) & \Xi^-(P+Q) \\ \Xi^+(P+Q) & G^-(P+Q)
\end{array} \right) \left( \begin{array}{cc} 0 & i \gamma_5 \\ i
\gamma_5 & 0 \end{array} \right) \\ & & \hspace{3cm} \left(
\begin{array}{cc} G^+(P) & \Xi^-(P) \\ \Xi^+(P) & G^-(P) \end{array}
\right) \left( \begin{array}{cc} 0 & i \gamma_5 \\ i \gamma_5 & 0
\end{array} \right) ] \\ 
&=& 1 - G \int \frac{d^4P}{(2
\pi)^4} \mbox{Tr}[ \Xi^-(P+Q) \gamma_5 \Xi^-(P) \gamma_5 + G^+(P+Q)
\gamma_5 G^-(P)\gamma_5 \\ & & \hspace{2.7cm} + G^-(P+Q) \gamma_5
G^+(P)\gamma_5+\Xi^+(P+Q) \gamma_5 \Xi^+(P) \gamma_5] \\ 
&=& 1+
G \int \frac{d^4P}{(2 \pi)^4} (1+{\bf \hat{p}} \cdot {\bf \widehat{p+q}})
\\ & & \times \frac{((p+q)_0+E_{\bf p+q})(p_0-E_{\bf
p})-\Delta^2+((p+q)_0-E_{\bf p+q})(p_0+E_{\bf
p})-\Delta^2}{[(p+q)_0^2-\xi_{\bf p+q}^2][p_0^2-\xi_{\bf p}^2]} \\ 
&=&
1+ 2 \: G \int \frac{d^3p}{(2
\pi)^3}(1+{\bf \hat{p}} \cdot {\bf \widehat{p+q}}) \\ & & \times\frac{1}{(2\xi_{\bf
p})(2\xi_{\bf p+q} )}\{\left[n(\xi_{\bf p})-n(\xi_{\bf
p+q})\right]\left(\xi_{\bf p+q} \xi_{\bf p} - E_{\bf p+q} E_{\bf p}
-\Delta^2\right) \\ & & \hspace{4.5cm} \left(\frac{1}{q_0+\xi_{\bf
p}-\xi_{\bf p+q}} - \frac{1}{q_0-\xi_{\bf p}+\xi_{\bf p+q}}\right) \\
& & \hspace{2.3cm}+\left[1-n(\xi_{\bf p})-n(\xi_{\bf
p+q})\right]\left(\xi_{\bf p+q} \xi_{\bf p} +E_{\bf p+q} E_{\bf p}
+\Delta^2\right) \\ & & \hspace{4.5cm} \left(\frac{1}{q_0-\xi_{\bf
p}-\xi_{\bf p+q}} - \frac{1}{q_0+\xi_{\bf p}+\xi_{\bf p+q}}\right) \}.
\end{eqnarray*}
This is Eq.~(\ref{rel_g_explicit}).

\section{Color-flavor traces in the polarization tensor for CFL}

In Appendix A we evaluated the polarization tensor for the
one-component, relativistic superfluid.  This involved traces over
Nambu-Gor'kov and Dirac indices, as well as a sum over Matsubara
frequencies.  Similar calculations are required to evaluate the
polarization tensor for CFL.  The two gap color-flavor structure of
the quark propagator complicates things.  We discuss in this appendix
the color-flavor traces that must be computed to evaluate the CFL
polarization tensor.  First we discuss the color-flavor traces
involved in the evaluation of the quasi-free response, and then we
discuss the color-flavor traces involved in the evaluation of the
collective response.

\subsection{Quasi-free response}

The contribution to the polarization tensor from the quasi-free
response of the medium comes from the first term on the right-hand
side of Fig.~\ref{nu-H_geo_Fig}.  Evaluating this diagram involves a
trace over Nambu-Gor'kov and Dirac indices, as well as a Matsubara sum
-- these calculations are similar to those in the relativistic case.
The complication in the present case comes from the trace over
color-flavor indices, due to the two gap structure of CFL.  The
color-flavor trace involves the quantities
$$ R^{(1)}_{mn} = \mbox{Tr}[c_V {\bf P_m} c_V {\bf P_n}] +
\mbox{Tr}[c_A {\bf P_m} c_A {\bf P_n}],$$
$$ R^{(2)}_{mn} = \mbox{Tr}[c_V {\bf P_m} c_A {\bf P_n}] +
\mbox{Tr}[c_A {\bf P_m} c_V {\bf P_n}],$$
$$ \tilde{R}^{(1)}_{mn} = \mbox{Tr}[c_V {\bf \tilde{P}_m} c_V {\bf
\tilde{P}_n}] + \mbox{Tr}[c_A {\bf \tilde{P}_m} c_A {\bf
\tilde{P}_n}],$$ and
$$ \tilde{R}^{(2)}_{mn} = \mbox{Tr}[c_V {\bf \tilde{P}_m} c_A {\bf
\tilde{P}_n}] + \mbox{Tr}[c_A {\bf \tilde{P}_m} c_V {\bf
\tilde{P}_n}].$$
Explicit values for these quantities are given in Eq.~(\ref{cf_free}).
\subsection{Collective response}
We now want to calculate the contribution to the quark polarization
tensor from the collective response associated with the $H$ boson, depicted in
Fig.~\ref{nu-H_geo_Fig}.  The color-flavor traces involved in
calculating this contribution are much simpler than those involved in
calculating the quasi-free response, since the $H$-quark coupling does
not mix the singlet and octet quasi-quarks.  To evaluate the numerator
of Fig.~\ref{nu-H_geo_Fig}, we only need the coefficients
$$ R^{(3)}_m = \mbox{Tr}[c_V {\bf P_m} M {\bf \tilde{P}_m}] =
\mbox{Tr}[c_V {\bf \tilde{P}_m} M {\bf P_m}] = \mbox{Tr}[c_A {\bf P_m}
M {\bf \tilde{P}_m}] = \mbox{Tr}[c_A {\bf \tilde{P}_m} M {\bf P_m}].
$$ This is explicitly evaluated in Eq.~(\ref{cf_I}).  To evaluate the
denominator, we only need the coefficients
$$ R^{(5)}_m = \mbox{Tr}[M {\bf P_m} M {\bf P_m}] = \mbox{Tr}[M {\bf
\tilde{P}_m} M {\bf \tilde{P}_m}].$$ This is explicitly evaluated in
Eq.~(\ref{cf_g}).

%\bibliography{cflbcsrpa} \bibliographystyle{h-physrev3.bst}

\end{document}